\documentclass{emulateapj}
\usepackage{natbib}

\newcommand{\Msol}{$M_{\odot}$}
\newcommand{\Zsol}{$Z_{\odot}$}
\newcommand{\Lsol}{$L_{\odot}$}

\newcommand{\etal}{\mbox{{\rm et~al.\ }}}
\newcommand{\Oxygen}{\mbox{${\rm {\rm ^{16}O}}$}}
\newcommand{\Carbon}{\mbox{${\rm {\rm ^{12}C}}$}}
\newcommand{\Fe}{\mbox{${\rm {\rm ^{56}Fe}}$}}
\newcommand{\Silicon}{\mbox{${\rm {\rm ^{28}Si}}$}}

\slugcomment{Submitted to ApJ on August 29; accepted on December 12, 2003}

\shorttitle{Can a standard IMF explain galaxy clusters?}
\shortauthors{Portinari et al.}
\begin{document}

\title{Can a ``standard'' Initial Mass Function explain \\
the metal enrichment in clusters of galaxies?}

\author{L.~Portinari\altaffilmark{1}, A.~Moretti\altaffilmark{2,3},
C.~Chiosi\altaffilmark{2} and J.~Sommer-Larsen\altaffilmark{1} }

\altaffiltext{1}{Theoretical Astrophysics Center, Juliane Maries Vej 30, 
DK-2100 Copenhagen \O, Denmark}
\altaffiltext{2}{Department of Astronomy, University of Padova, 
Vicolo dell'Osservatorio 2, I-35122 Padova, Italy}
\altaffiltext{3}{INAF, Observatory of Padova, 
Vicolo dell'Osservatorio 5, I-35122 Padova, Italy}
\email{lportina,jslarsen@tac.dk; moretti,chiosi@pd.astro.it}

\begin{abstract}
\noindent
It is frequently debated in literature whether a ``standard'' Initial Mass 
Function (IMF) --- meaning an IMF of the kind usually adopted to explain 
the chemical evolution in the local Solar Neighbourhood --- can account for 
the observed metal enrichment and Iron Mass--to--Light Ratio in clusters 
of galaxies. We address this problem by means of straightforward
estimates that should hold independently of the details of chemical
evolution models. 

It is crucial to compute self--consistently the amount of mass 
and metals locked--up in stars, by accounting for the stellar 
mass--to--light ratio predicted by a given IMF.
It becomes then clear that a ``standard'' Solar Neighbourhood IMF 
cannot provide enough metals to account for the observed chemical properties 
in clusters:
clusters of galaxies and the local environment must be characterized by 
different IMFs. 

Alternatively, if we require the IMF to be universal, 
in order to explain clusters such an IMF must be much 
more efficient in metal production than usually estimated for the Solar 
Vicinity. In this case, substantial loss of metals is required from the Solar 
Neighbourhood and from disc galaxies in general. This ``non--standard'' 
scenario of the local
chemical evolution would challenge our present understanding of the Milky Way 
and of disc galaxy formation.
\end{abstract}

\keywords{Galaxies: chemical evolution; clusters; intra--cluster medium;
stars: Initial Mass Function}
\section{Introduction}

\noindent
There seems to be no consensus in literature as to whether a 
``standard'' Initial Mass Function (IMF) --- by which term we indicate 
an IMF of the kind generally adopted to explain the chemical properties 
of the Solar Vicinity, in the standard models with infall and no outflows --- 
can account for the observed level of metal enrichment and Iron 
Mass--to--Light Ratio
\citep{Ci91,R93}
in clusters of galaxies. 

The efficiency of metal enrichment from a stellar population
is estimated by computing the global (or net) yield 
\citep{T80,Pag97}:
\begin{equation}
\label{eq:yield}
y = \frac{1}{1-R} \, \int_{M_{TO}(t)}^{M_s} p_Z(M) \Phi(M) \, dM
\end{equation}
where $M_{TO}(t)$ is the turn--off stellar mass of lifetime corresponding 
to the age $t$
of the population, $M_s$ is the upper mass limit of the IMF, $p_Z(M)$ 
is the mass fraction of newly synthesized metals
ejected by a star of mass $M$ and $\Phi(M)$ is the IMF. Hence the integral 
expresses the amount 
of metals globally produced by a stellar generation up to its present age.
$R$ represents the ``returned fraction'' of gas re--ejected by dying stars, 
so that $1-R$ is the fraction of mass locked up into living stars
and remnants.
The definition of the net yield\footnote{In this paper, the term ``yield''
will generally indicate the net yield $y$ of a stellar population as defined
in Eq.~\ref{eq:yield}; we will explicitly speak, instead, of ``stellar yields''
to indicate the $p_Z(M)$ resulting from nucleosynthesis calculations.}
remarks that the efficiency of metal enrichment
depends not just on the amount of metals produced per
mass involved in star formation, but on the ratio between this and
the mass that remains locked in stars. The locked--up fraction, related to 
the number of ever--lived low--mass stars in the IMF, is as crucial to the 
overall enrichment as is the number of the high mass stars directly responsible
for the production of metals. By ``ever--lived stars'' we mean stars with 
lifetimes longer than a Hubble time, typically $M<$0.9~\Msol.

As pointed out by \cite{A92},
it would be surprising if the same ``recipe'' could account
at the same time for the Solar Vicinity and for galaxy clusters,
since the observed yield
is definitely very different in the two environments. In 
the Solar Vicinity the gas mass is $\sim$20\% of the stellar 
mass and its metallicity is $\sim$\Zsol, while in clusters of galaxies 
the intra--cluster gas mass is estimated to be 5--10 times the stellar mass
\citep{A92,Lin03}
and its typical metallicity is $\sim$0.3~\Zsol.
The stellar metallicity, on the other hand, is comparable (roughly solar) 
in the two environments. This na{\"\i}ve estimate outlines the  difference 
between clusters and the Solar Vicinity :
%
\begin{equation}
\label{eq:yield_SV}
y_{SV} \sim \frac{Z_{\odot} \times M_* + Z_{\odot} \times (0.2 \, M_*)}{M_*}
 = 1.2 \, Z_{\odot}
\end{equation}
\begin{equation}
\label{eq:yield_cl}
y_{cl} \sim \frac{Z_{\odot} \times M_* + 0.3 \, Z_{\odot} \times 
(5-10 \, M_*)}{M_*} = 2.5-4 \, Z_{\odot}
\end{equation}
\citep[see also][]{Pag02}.
The difference is possibly larger than this, since the typical metallicity
of the stellar populations in clusters may be higher than the local one:
galaxies in clusters are in fact dominated in mass by the bright ellipticals 
with metallicities between solar and supersolar,
while in the Solar Vicinity the metallicity distribution peaks around 
--0.2~dex.
We assess in detail the observed yield in the Solar Neighbourhood 
versus cluster in the Appendix, distinguishing iron from $\alpha$--elements.
All in all, the observed yield in clusters is 
3--4 times larger than in the Solar Vicinity.

Gaseous inflows and outflows can alter the ``apparent'' or ``effective yield'',
as deduced from the observed metallicities, 
with respect to the ``true'' net yield of the underlying stellar 
population as defined in Eq.~\ref{eq:yield} \citep{Ed90, EG95, Pag97}.
However, it is hard to imagine gas flows to be so prominent on the scale 
of clusters as to substantially modify the apparent yield over so large scales
and masses.
It might be somewhat easier, though still a complex issue, to invoke these 
effects 
on the scale of individual galaxies and of the Solar Vicinity;
see our conclusions in \S\ref{sect:concl_outflows}. The effect of flows 
is in fact, in most cases, to decrease the apparent yield below the true yield.

As a consequence of the strikingly different observed yield, many authors 
resorted to 
``non--standard'' IMFs to explain the level of metal enrichment in clusters,
invoking in turn top--heavy IMFs \citep{DFJ91b, MG95, GMl97, GM97, LM96},
bimodal IMFs \citep{A92, EA95}, variable IMFs \citep{C2000, MPC03, Fin03}
or contribution from Population~III hypernov\ae\ \citep{Loe2001, Baum03}.

On the other hand, the arguments brought in favour of an invariant, standard 
IMF are {\bf (i)} that the Iron Mass--to--Light Ratio (IMLR) observed in 
clusters is compatible with the predictions from the Salpeter IMF 
\citep{R93, R97,R2003} or {\bf (ii)} that the [$\alpha$/Fe] abundance ratios 
in the intra--cluster medium (ICM) are compatible, within uncertainties, with 
those in the Sun and in local disc stars \citep{IA97, Wy97, R97, R2003}.
However, neither of these two diagnostics
is sensitive to the amount of mass (and metals!) locked--up in low-mass stars.
Very low-mass, ever lived stars do not contribute metals nor luminosity, 
but they act as an effective {\it sink} of metals. 
This effect is crucial to determine the partition of metals between
the stars and the ICM.

{\bf (i)} Let's define the typical IMLR expected from a stellar population with
a given IMF as:
\[ IMLR_{SSP} = \frac{MFe}{L_B} \]
where $L_B$ is the luminosity of a Single Stellar Population (SSP)
of initial mass 1~\Msol\ and $MFe$ is the iron mass produced by the same SSP.
Consider, as an example, two Salpeter IMFs differing only
in the low--mass limit, say [0.1--100]~\Msol\ and [0.5--100]~\Msol\ 
respectively:
these two IMFs have the same typical IMLR$_{SSP}$, because stars between 
{\mbox{0.1--0.5~\Msol}} do not contribute neither metals nor substantial
luminosity.
However these IMFs imply, 
{\it for the same IMLR$_{SSP}$ and observed stellar metallicity}, 
very different amounts of mass and metals locked up in the stellar component, 
correspondingly changing
the remaining fraction of the produced metals available to enrich the ICM.

{\bf (ii)} Similarly, [$\alpha$/Fe] abundance ratios trace the {\it relative} 
proportion of type~II and type~Ia supernov\ae\ (SN~II and SN~Ia), 
that is to say about the relative number of massive 
stars with respect to intermediate (down to $\sim$ solar) mass stars.
This is sensitive to
the shape and slope of the IMF above $\sim$1~\Msol, but it is insensitive to
the amount of mass locked in ever--lived stars of lower mass 
\citep[as acknowledged in fact by][]{Wy97} 
or to the {\it global} amount of 
metals produced.

The amount of mass locked in low mass, ever--lived stars is strictly 
related to the stellar mass--to--light (M/L) ratio predicted by a given IMF 
\citep{R93}. In this paper we stress
the importance of taking into account self--consistently the stellar M/L ratio,
in order to understand whether a standard IMF can in fact explain the observed
metal enrichment of clusters of galaxies. 

\begin{figure*}
\epsscale{0.77}
\plotone{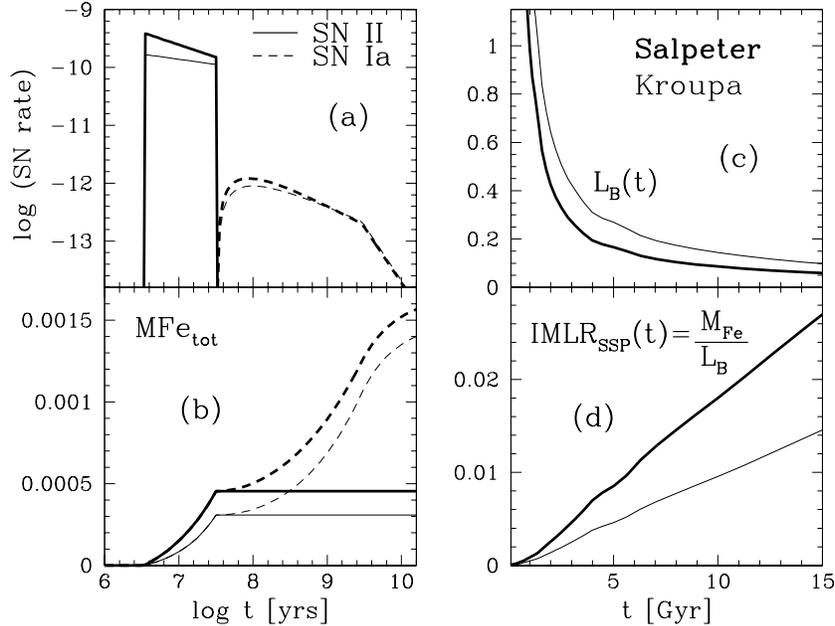}
\caption{(a) Evolution of the rate of SN~II and SN~Ia, in number per year, for 
a Salpeter and a Kroupa SSP of 1~\Msol\ (thick and thin lines, respectively).
(b) Corresponding cumulative iron production from SN~II and SN~Ia, in~\Msol. 
(c) B--band luminosity evolution of a Salpeter and a Kroupa SSP of 1~\Msol,
in $L_{B, \odot}$. 
(d) Evolution of the characteristic IMLR$_{SSP}$ of a Salpeter and 
a Kroupa SSP.
\label{fig:rateSN_MFe_LB_IMLRssp} }.
\end{figure*}

Our approach is very straightforward: for any given IMF, one can compute
the corresponding rates of SN~II and SN~Ia, and the rate
of production of iron $MFe_{tot}(t)$ (Fig.~\ref{fig:rateSN_MFe_LB_IMLRssp}ab) 
--- or of any other element.
For the same IMF, the corresponding SSP derived from stellar isochrones gives 
the luminosity evolution $L_B(t)$ (Fig.~\ref{fig:rateSN_MFe_LB_IMLRssp}c), 
and the ratio 
$MFe_{tot}(t)/L_B(t)$ gives the typical IMLR$_{SSP}(t)$ expected from the 
stellar population, as a function of its age 
(Fig.~\ref{fig:rateSN_MFe_LB_IMLRssp}d). Of course,
IMLR$_{SSP}$ is not the IMLR observed in the ICM, 
since part of the metals produced will be ``eaten up'' by successive stellar
populations to build up the observed stellar metallicities in cluster galaxies.
Therefore, our second step is to consider the mass in stars and the fraction 
of metals locked in stars 
{\it consistent} with the adopted IMF and with the observed stellar 
metallicities; only what remains out of the global metal production, is 
in principle available to enrich the ICM and can be compared to the observed 
ICM metallicity.

We will discuss two cases of IMF:
the Salpeter IMF (mostly for comparison and discussion of previous literature;
Section~\ref{sect:Salpeter}) and the Kroupa IMF (an example of ``standard'' 
IMF for the Solar Neighbourhood; Section~\ref{sect:Kroupa}). In 
Section~\ref{sect:discussion} we will discuss our results, that clearly 
indicate that a Salpeter IMF --- and even less a standard Solar Neighbourhood 
IMF --- is unable to account for the level of metal enrichment in clusters 
of galaxies; these results are robust with respect to uncertainties concerning
the theoretical stellar yields, the rate of SN~Ia and the details in the
star formation history of cluster galaxies. We also outline the qualitative 
features of an IMF that should be able to reproduce clusters of galaxies. 
Finally, in 
Section~\ref{sect:conclusions} we conclude that we are left either with
a non--universal IMF (different between clusters and Solar Neighbourhood), 
or with a non--standard scenario for the chemical evolution of the Solar 
Neighbourhood; we discuss each possibility in turn.

\section{Predictions from the Salpeter IMF}
\label{sect:Salpeter}

\noindent
We will first discuss the most ``classic'' case, the \cite{Sa55} IMF
with constant power--law slope and mass limits [0.1--100]~\Msol;
all the computations in this Section~\ref{sect:Salpeter}, as well as the terms
``Salpeter IMF'' or ``Salpeter SSP'' hereinafter, refer to the above mentioned
mass limits.

We remark, though, that the Salpeter IMF is {\it not}
a ``standard'' IMF in the sense explained in the introduction: it is 
{\it too efficient} in metal enrichment, i.e.\ its typical global yield is too
high, to match the typical abundances in the Solar Neighbourhood and in disc 
galaxies, 
within standard chemical evolution models \citep{Tsu97,TGB98,Gra2000,PSLT03}.
Besides, from observed star counts in the Solar Neighbourhood and arguments
related to the stellar M/L ratio in disc galaxies, there is by now a general
consensus that the IMF is not a single--slope power-law extrapolated 
down to 0.1~\Msol, but it ``bends over'' below $\sim$1~\Msol\ implying a 
smaller percentage of low--mass stars 
\citep[][and references therein]{Kr2001,Kr2002,Cha2003,PSLT03}.

Nevertheless, the ``classic'' Salpeter IMF is still very widely adopted 
in literature for chemical and spectro--photometric models, and it has been
often adopted to model clusters of galaxies, so it is
mandatory to consider it for the sake of comparison and comment
of literature results.

\begin{deluxetable*}{c c c c c c c c c c c}
\tablecolumns{11}
\tablewidth{0pt}
\tablecaption{ {\mbox{Evolution of the metal production from
a Salpeter SSP}} \label{tab1}}
\tablehead{ 
\colhead{$^{(1)}$} & \colhead{$^{(2)}$} & \colhead{$^{(3)}$} & 
\colhead{$^{(4)}$} & \colhead{$^{(5)}$} & \colhead{$^{(6)}$} & 
\colhead{$^{(7)}$} & \colhead{$^{(8)}$} & \colhead{$^{(9)}$} & 
\colhead{$^{(10)}$} & \colhead{$^{(11)}$} \\
\colhead{t[Gyr]} & \colhead{M$_{TO}$(t)} & \colhead{$N_{Ia}$} & 
\colhead{$\frac{N_{Ia}}{N_{II}}$} & \colhead{$MFe_{Ia}$} & 
\colhead{$\frac{MFe_{Ia}}{MFe_{II}}$} & \colhead{$MFe_{tot}$} & 
\colhead{[O/Fe]} & \colhead{$MSi_{Ia}$} & \colhead{$MSi_{tot}$} & 
\colhead{[Si/Fe]}
}
\startdata
  1 & 2.2 & 5.5e-4  & 0.09 & 3.9e-4 & 0.8 & 8.7e-4 &  +0.22 & 8.5e-5 & 8.8e-4 & +0.25 \\
  2 & 1.6 & 1.0e-3  & 0.16 & 7.0e-4 & 1.5 & 1.2e-3 &  +0.09 & 1.5e-4 & 9.5e-4 & +0.14 \\
  5 & 1.2 & 1.4e-3  & 0.22 & 9.9e-4 & 2.0 & 1.5e-3 & --0.01 & 2.2e-4 & 1.0e-3 & +0.06 \\
 10 & 1.0 & 1.55e-3 & 0.24 & 1.1e-3 & 2.3 & 1.6e-3 & --0.04 & 2.4e-4 & 1.0e-3 & +0.06 \\
 15 & 0.9 & 1.6e-3  & 0.25 & 1.1e-3 & 2.3 & 1.6e-3 & --0.04 & 2.5e-4 & 1.0e-3 & +0.06 \\
\enddata
\tablecomments{
(1) Age of the SSP in Gyr.
(2) Corresponding indicative turn-off mass.
(3) Cumulative number of SN~Ia exploded up to age $t$.
(4) Relative number of SN~Ia vs.\ SN~II exploded up to age $t$.
(5) Cumulative iron mass produced by SN~Ia.
(6) Ratio of iron masses produced by SN~Ia and SN~II.
(7) Total iron mass produced up to age $t$ by SN~II+SN~Ia.
(8) [O/Fe] ratio of the cumulative metal production of the SSP.
(9) Cumulative silicon mass produced by SN~Ia.
(10) Total silicon mass produced up to age $t$ by SN~II+SN~Ia.
(11) [Si/Fe] ratio of the cumulative metal production of the SSP.
 }
\end{deluxetable*}

\subsection{Metal production}
\label{sect:metals_salp_ssp}

\noindent
Let us compute the amount of newly synthesized metals expected to be
released by a Salpeter SSP of initial mass 1~\Msol.
Adopting the chemical yields by \cite{PCB98}, 
a Salpeter SSP produces 
\[ MO_{II}=0.01~M_{\odot} \] 
\[ MFe_{II}=4.8 \times 10^{-4}~M_{\odot} \]
of new \Oxygen\ and \Fe\ from SN~II. This estimate 
takes into account that the original SN models by \cite{WW95}, 
which were ultimately the base 
for the Portinari \etal yields, are 
too efficient in iron production.
Here we decreased the iron yield of SN~II by a factor of 1.5, so that
the typical [O/Fe] ratio of SN~II ejecta is [O/Fe]$\sim$+0.5 as indicated by 
abundances in halo stars.\footnote{We adopt
as solar abundances [O/H]$_{\odot}$=8.87, [Si/H]$_{\odot}$=7.55 and 
[Fe/H]$_{\odot}$=7.5 \citep{Grev96}.
Notice that the recent photospheric abundance
of iron is in excellent agreement with the meteoritic value \citep{Grev99}.
These abundances correspond to mass fractions of $X_O=8.3 \times 10^{-3}$, 
$X_{Si}=6.9 \times 10^{-4}$ and $X_{Fe}=1.2 \times 10^{-3}$ respectively, 
i.e.\ to an oxygen mass 7 times larger than the iron mass in the Sun.
\label{foot:sol_ab}}
This amount of iron production is in close agreement with the following
``rule of thumb'' estimate: with a Salpeter IMF, a 1~\Msol\ SSP results in
$7.4 \times 10^{-3}$~SN~II ($M>8$~\Msol),
and each of these produces on average 0.07~\Msol\ of \Fe\
(like SN~1987a). We neglect here the metallicity dependence of stellar yields,
which for oxygen and iron becomes relevant only for $Z>$~\Zsol.

The release of new oxygen and iron by SN~II
can be considered instantaneous for the purpose of this paper. 
The iron from SN~Ia is released
instead over longer timescales (Fig.~\ref{fig:rateSN_MFe_LB_IMLRssp}ab). 
We adopt the SN~Ia rate formalism by \cite{GR83},
with the classic mass limits [3--16]~\Msol\ and frequency coefficient $A=0.07$
for the binary systems progenitors of SN~Ia; this calibration is suited 
to reproduce the 
typical SN~II/SN~Ia ratio in Milky Way--type galaxies and the evolution of the
abundance ratios in the Solar Neighbourhood \citep{CMG97,PCB98}.

\begin{figure}
\epsscale{0.98}
\plotone{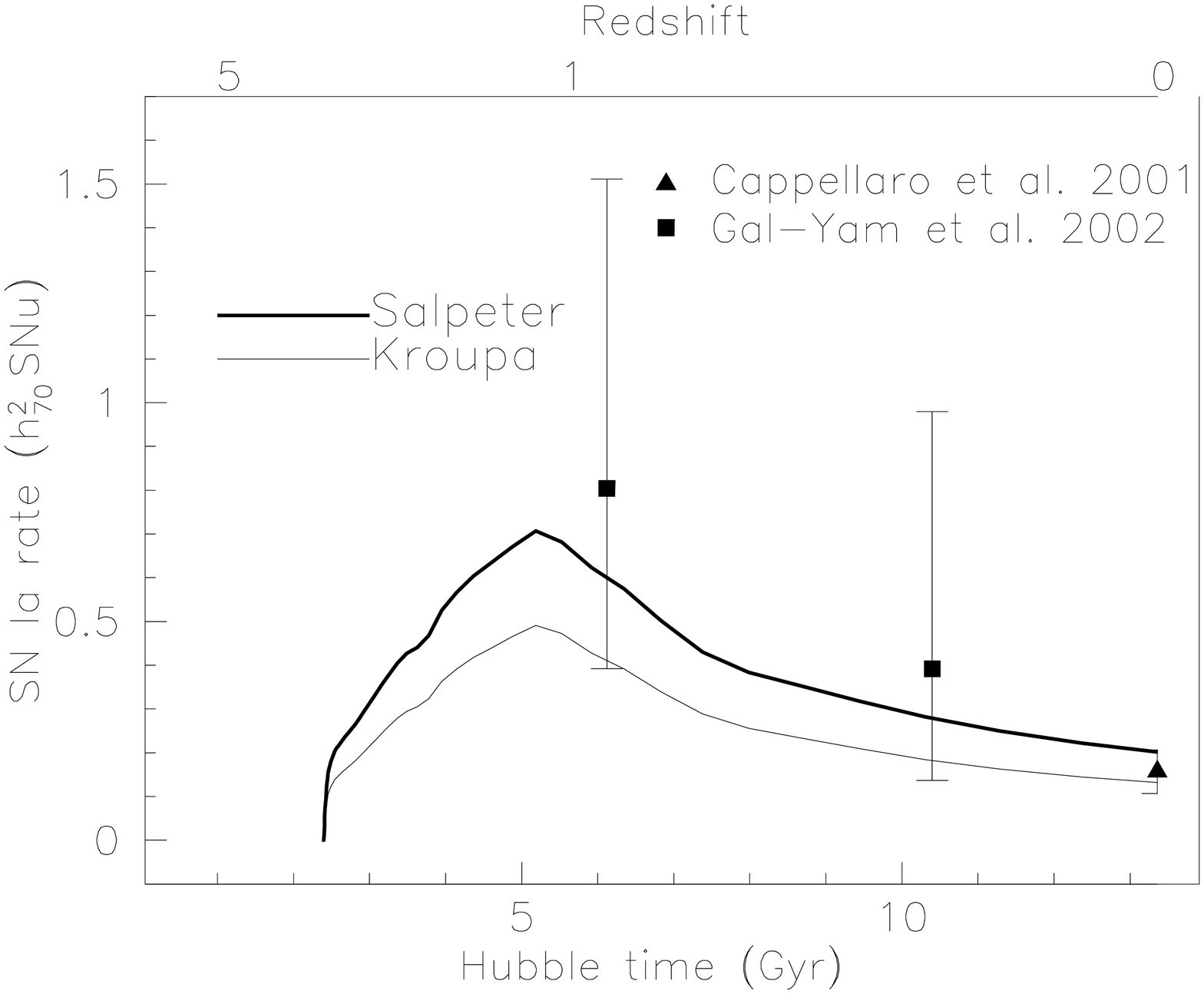}
\caption{Evolution of the SN~Ia rate (in SNu) with respect to Hubble time,
for a Salpeter and a Kroupa SSP (thick  and thin lines, respectively)
of 11 Gyr of age. Data from \citep{Capp01} and \citep{galyam02}
\label{fig:SNIa_rate_dat} }.
\end{figure}

In Fig.~\ref{fig:SNIa_rate_dat}, we plot the SN~Ia rate for our SSPs in SNu
(SN units: number of SN\ae\ per 100~yrs per 10$^{10} \, L_{B,\odot}$) 
and compare it to observational estimates. The rate 
in SNu is obtained as the ratio of the SN~Ia rate of the SSP
(Fig.~\ref{fig:rateSN_MFe_LB_IMLRssp}a) versus the corresponding luminosity
evolution $L_B(t)$
(Fig.~\ref{fig:rateSN_MFe_LB_IMLRssp}c). Just like the IMLR$_{SSP}$ or the
relative number of SN~II vs.\ SN~Ia, the SN rate in SNu is independent 
of the amount of mass locked in very low mass stars: it is only sensitive
to the shape of the IMF above about 1~\Msol, whence both SN\ae\ and luminosity
originate. We compare our SSP predictions to the local SN rate
in elliptical galaxies \citep{Capp01}, where the stellar population, with
an early and short star formation history, more closely resembles a SSP.
The local SN rate in ellipticals is very well matched by our models, which
is remarkable since our SN~Ia prescription is suited and calibrated for the
Solar Neighbourhood.

At redshift $z>0$, we compare our predictions to the results from an HST
high--redshift search for SN~Ia in clusters \citep{galyam02, Maoz03} --- thus
mostly sampling elliptical galaxies. The agreement is good
also at higher redshift,
considering that we are adopting a ``standard Solar Neighbourhood'' SN~Ia 
prescription.
In Fig.~\ref{fig:SNIa_rate_dat},  we assumed an age of 11~Gyr for the SSP;
but the agreement with the data is acceptable for SSP ages between 10--12~Gyr.

We compute and list in Table~\ref{tab1} the cumulative number $N_{Ia}$ of SN~Ia
in a Salpeter SSP as a function of time,
and the related iron production $MFe_{Ia}$.
We assume that each SN~Ia produces 0.7~\Msol\ of iron.
By the end of the evolution, the number of SN~Ia is about 
1/4 of that of SN~II and the iron contribution is 1/3 from SN~II and 2/3 
from SN~Ia (see also Fig.~\ref{fig:rateSN_MFe_LB_IMLRssp}b).
These values are compatible with the solar proportions, but we stress
that these represent {\it relative} proportions while the {\it global}
metal production depends on the mass limits of the IMF, i.e.\ its 
normalization.

The oxygen production from SN~Ia is minor: each SN~Ia produces 
0.143~\Msol\ of oxygen \citep{Iwa99}
so globally, for a Salpeter SSP,
SN~Ia altogether contribute 
\[ MO_{Ia}=2 \times 10^{-4}~M_{\odot}, \] 
absolutely negligible
with respect to the SN~II production. In column (8) of Table~\ref{tab1} 
we list the typical [O/Fe] ratio of the global metal production (SN~II+SN~Ia)
of the SSP as a function of time. For $t >$5~Gyr the [O/Fe] ratio
is roughly solar (see also Fig.~\ref{fig:partOMLR}b, solid line).

\subsection{Luminosity, stellar mass and IMLR}

\noindent
We now derive the typical IMLR$_{SSP}$ by combining the global iron 
production $MFe_{tot}(t)$ in Table~\ref{tab1} and 
Fig.~\ref{fig:rateSN_MFe_LB_IMLRssp}b, with the luminosity evolution 
of a Salpeter SSP.
In Fig.~\ref{fig:rateSN_MFe_LB_IMLRssp}c we show the luminosity evolution 
for a Salpeter IMF 
computed from the isochrones by \cite{Gir02} with solar metallicity and 
${\cal M}_{B \odot}$=5.489; in Fig.~\ref{fig:rateSN_MFe_LB_IMLRssp}d 
we show the corresponding 
evolution of the characteristic IMLR$_{SSP}=MFe_{tot}/L_B$. The values of
$L_B$ and IMLR$_{SSP}$ for some representative ages are listed in 
Table~\ref{tab2}. 

The observed IMLR in the ICM is IMLR$_{ICM} \sim 0.01-0.015$~\Msol/\Lsol\
\citep{F2000,Fin01,DeGra03}.\footnote{We note that
the iron abundances measured in the ICM are often expressed in terms
of the old photospheric value for the solar iron abundance by \cite{AG89}.
More recent photospheric values agree instead with the meteoritic value
which is lower by $\sim$0.17~dex, i.e.\ a factor of 1.5 in mass/number 
abundance (see also 
footnote~\ref{foot:sol_ab}) . This has induced a great deal
of confusion on the interpretation of the abundance ratios in the ICM
\citep{IA97,Wy97}. While the latest X--ray studies have started to refer
consistently to the updated solar iron abundance \citep{Baum03,DeGra03},
we notice that fortunately the reference solar value adopted in previous 
years bears no effects on the estimated iron mass or IMLR in the ICM.
In fact, in the spectral fits used to infer elemental abundances from X--ray 
lines, it is the {\it absolute abundance} by number of the coolants --- 
iron ions 
in this case --- present in the chemical mixture, that ultimately determines
the line strength to be compared to the observed one. 
Whether this absolute abundance is expressed as a fraction of the old or
of the updated solar abundance is a secondary issue from the point of view
of the {\it global} iron mass MFe$_{ICM}$.
Henceforth, although attention must be paid
to the adopted solar values when comparing e.g.\ ICM data to 
Galactic data, fortunately the estimates of global MFe$_{ICM}$ and 
IMLR$_{ICM}$ are robust absolute values and do not scale with different 
assumptions for the solar abundance \citep[see also][]{Baum03}.}
From Table~\ref{tab2} and Fig.~\ref{fig:rateSN_MFe_LB_IMLRssp}d, 
we can anticipate that the 
IMLR$_{SSP}$ of a Salpeter SSP {\it can} be compatible with the IMLR$_{ICM}$
observed in the ICM {\it provided} we are dealing with old
stellar populations ($t \geq 10$~Gyr) and {\it provided} a large part of the
iron gets dispersed into the ICM.

Notice from Fig.~\ref{fig:rateSN_MFe_LB_IMLRssp}b that the bulk of the iron 
production (70--80\%) from the SSP takes place in fact within the first 
2--3~Gyr. 
An early iron production from SN~Ia is not only in agreement
with the empirical  estimates of the SN~Ia rate in clusters up to redshift
$z  \sim$1  (Fig.~\ref{fig:SNIa_rate_dat}) but also, if  SN~Ia
contribute a significant fraction of the iron in the  ICM, with
the fact that the observed ICM iron abundance is invariant in clusters
up to $z \sim$1.2  \citep{Tozzi03}. 
The evolution of the IMLR$_{SSP}$ at late times is then dominated 
by the  fading luminosity rather than by late iron production
(Fig.~\ref{fig:rateSN_MFe_LB_IMLRssp}cd).  Therefore, the  age  of the
stellar population is as crucial  to the IMLR$_{SSP}$ as the amount of
iron produced; the presence of a somewhat younger stellar component in
the  cluster,  for  instance  due  to spiral  galaxies  or  late  star
formation  episodes, can  significantly  reduce the  IMLR  due to  the
increased luminosity (cf.\ IMLR$_{SSP}<$0.01 at $t \leq$5~Gyr.)

To ascertain whether the Salpeter IMF can enrich the ICM to the observed 
levels, it is now crucial to estimate what fraction of the produced metals 
is effectively shed into the ICM versus the fraction stored in the
stellar component. 

\begin{deluxetable}{ccccccc}
\tablecolumns{7}
\tablewidth{0pt}
\tablecaption{Evolution of luminosity, M/L ratio and IMLR
{\mbox{for a Salpeter SSP}} \label{tab2}}
\tablehead{
\colhead{$^{(1)}$} & \colhead{$^{(2)}$} & \colhead{$^{(3)}$} & 
\colhead{$^{(4)}$} & \colhead{$^{(5)}$} & \colhead{$^{(6)}$} & 
\colhead{$^{(7)}$} \\
\colhead{t[Gyr]} & \colhead{${\cal M}_B$} & \colhead{$L_B$[\Lsol]} & 
\colhead{$M_*$} & \colhead{$\frac{M_*}{L_B}$} & \colhead{IMLR$_{SSP}$} & 
\colhead{SiMLR$_{SSP}$} 
}
\startdata
  1 & 5.506 & 0.984 & 0.78 & 0.80 & 8.8e-4 & 8.9e-4 \\
  2 & 6.416 & 0.426 & 0.75 & 1.74 & 2.8e-3 & 2.2e-3 \\
  5 & 7.435 & 0.167 & 0.72 & 4.31 & 9.0e-3 & 6.0e-3 \\
 10 & 8.145 & 0.087 & 0.71 & 8.16 & 1.8e-2 & 1.1e-2 \\
 15 & 8.554 & 0.059 & 0.70 & 12.5 & 2.7e-2 & 1.7e-2 \\
\enddata
\tablecomments{
(1) Age of the SSP in Gyr.
(2) B--band  magnitude of the SSP.
(3) B--band luminosity of the SSP (${\cal M}_{B \odot}$=5.489).
(4) Mass fraction of the initial 1~\Msol\ SSP remaining in stars and remnants.
(5) M/L ratio of the stellar(+remnants) component of the SSP.
(6) Global IMLR of the SSP, computed from MFe$_{tot}(t)$ in Table~\ref{tab1}.
(7) Global SiMLR of the SSP, computed from $MSi_{tot}(t)$ in Table~\ref{tab1}.
}
\end{deluxetable}

\begin{figure}
\plotone{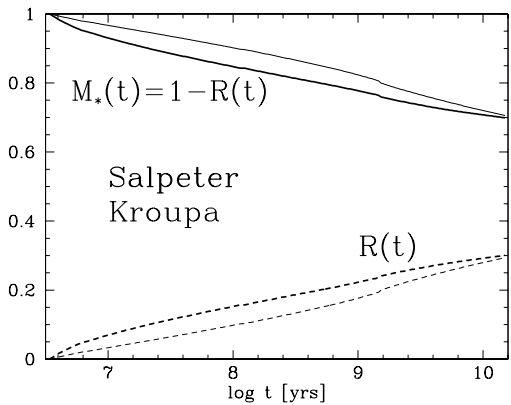}
\caption{Evolution of the returned fraction $R$ and of the complementary 
locked--up fraction $M_*$ for a Salpeter and a Kroupa SSP (thick and thin 
lines, respectively). Fractions are with respect to the initial mass of the 
SSP.
\label{fig:returned} }.
\end{figure}

\subsection{The metal partition between stars and ICM}
\label{sect:partition}

\noindent
We estimate the amount of metals locked in the stellar component
--- including living stars and remnants --- simply as:
%
\begin{equation}
\label{eq:MFe*}
{\rm MFe}_*(t) = X_{Fe,*} \times M_*(t)
\end{equation}
where $X_{Fe,*}$ is the observed metallicity in the stars
and $M_*(t)$ is the mass in stars predicted {\it consistently} from the SSP 
evolution (Table~\ref{tab2} and Fig.~\ref{fig:returned}). As a consequence, 
the remaining iron mass 
available to be dispersed in the ICM is:
\begin{equation}
\label{eq:diffMFe}
{\rm MFe}_{ICM}(t) = MFe_{tot}(t) - {\rm MFe}_*(t)
\end{equation}
where $MFe_{tot}(t)$ is the iron mass globally produced by the SSP, from 
Fig.~\ref{fig:rateSN_MFe_LB_IMLRssp}b and Table~\ref{tab1}. Analogous 
calculations apply for the partition of other chemical elements.

In Eq.~\ref{eq:diffMFe}, it may at first look awkward 
to consider the metals ${\rm MFe}_*$ locked in the stars of a SSP, 
as a fraction of the metals $MFe_{tot}$ produced by the same SSP:
a SSP is a burst of star formation formed out of gas with some 
pre-existing metallicity, and only later in time it releases
newly synthesized metals; a SSP cannot obviously enrich itself. 
In real galaxies, the stellar population is not a SSP but the result 
of a complex star formation history extended in time, even if fast 
and intensive as in elliptical galaxies. Subsequent populations 
lock up part of the metals produced by the previous ones, in a way that 
depends non--trivially on the detailed star formation, inflow and outflow 
history \citep{TC2002}. Nevertheless, SSP--based computations as in 
Tables~\ref{tab1} and~\ref{tab2} give a straightforward estimate 
of the global amount of metals produced by the stellar populations
in the galaxies, and {\it somehow} a fraction of those metals has been 
recycled and locked in the stars to reach the observed stellar metallicities.
For the purpose of the simple estimates in this paper,
the complex star formation and enrichment history of ellipticals
can be schematically treated as a ``self--enriched SSP''.

In Fig.~\ref{fig:returned} we show the evolution of the fraction $M_*=1-R$ 
of the
initial mass of the SSP that is locked up in living stars and remnants
(see Eq.~\ref{eq:yield}); $M_*$ decreases in time 
because of gas re-ejection by dying stars over long timescales. In 
Table~\ref{tab2} we list $M_*$ and the corresponding stellar M/L ratio,
for some representative ages.
The locked--up
fraction $M_*$ is derived from the remnant masses by \cite{PCB98}
for massive stars, and by \cite{Mar01} for low and intermediate mass
stars, hence the computation is self--consistent with the luminosity evolution
from the isochrones by \cite{Gir02}: everything is based on the same set
of stellar models from the Padua group. 

\begin{figure}
\plotone{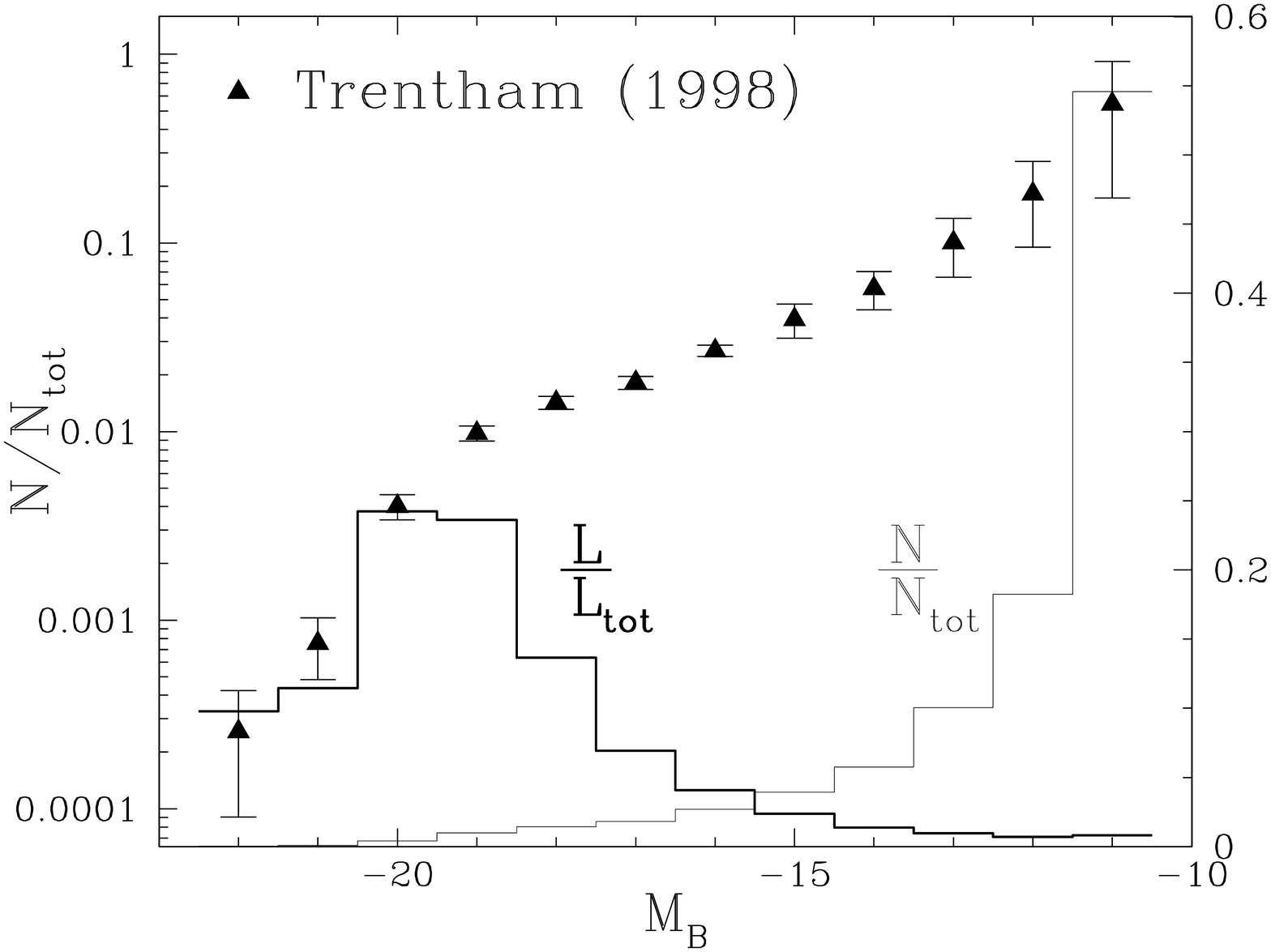}
\caption{Observed cluster luminosity function by \citet[][triangles and
left vertical axis]{Tr98}, and corresponding contribution of the different
luminosity bins to the number of galaxies and to the global luminosity
(right axis, in linear scale). Dwarf galaxies largely dominate in number, but
the bulk of the luminosity (and therefore stellar mass) is in massive
galaxies with $L \sim L_*$.
\label{fig:LF} }.
\end{figure}

We need now an estimate of the typical metallicities of the stellar component
in Eq.~\ref{eq:MFe*}. Although dwarf galaxies largely dominate in number,
the light, stellar mass and metal production in clusters 
are dominated by the massive ellipticals 
\citep[Fig.~\ref{fig:LF} and][]{Tho99,MPC03}. These objects have a typical 
metallicity that is +0.2 dex in the central regions and about solar 
at the effective radius, with an overall uniform 
[$\alpha$/Fe]=+0.2 \citep{Ar97, JorI99, Trag2000a, Trag2000b, Mehl03}.

\begin{deluxetable*}{c c c c c c c c c c c}
\tabletypesize{\scriptsize}
\tablecolumns{11}
\tablewidth{0pt}
\tablecaption{ {\mbox{Partition of metals between stars and ICM
with the Salpeter IMF (Case A)}} \label{tab3} }
\tablehead{
\colhead{$^{(1)}$} & \colhead{$^{(2)}$} & \colhead{$^{(3)}$} & 
\colhead{$^{(4)}$} & \colhead{$^{(5)}$} & \colhead{$^{(6)}$} & 
\colhead{$^{(7)}$} & \colhead{$^{(8)}$} & \colhead{$^{(9)}$} &
\colhead{$^{(10)}$} & \colhead{$^{(11)}$} \\
\colhead{t[Gyr]} & \colhead{MFe$_*$} & \colhead{MFe$_{ICM}$} & 
\colhead{IMLR$_*$} & \colhead{IMLR$_{ICM}$} & \colhead{MO$_*$} & 
\colhead{MO$_{ICM}$} & \colhead{[O/Fe]$_{ICM}$} & \colhead{MSi$_*$} & 
\colhead{MSi$_{ICM}$} & \colhead{SiMLR$_{ICM}$} 
}
\startdata
  1 & 9.4e-4 &   ---  & 9.6e-4 &   ---  & 1.0e-2 & ---  &   ---  & 8.6e-4 & 2.8e-5 & 2.8e-5 \\
  2 & 9.0e-4 & 3.0e-4 & 2.1e-3 & 7.0e-4 & 1.0e-2 & ---  &   ---  & 8.2e-4 & 1.3e-4 & 3.1e-4 \\
  5 & 8.6e-4 & 6.4e-4 & 5.0e-3 & 3.8e-3 & 9.6e-3 & 4e-4 & --1.05 & 7.8e-4 & 2.3e-4 & 1.4e-3 \\
 10 & 8.5e-4 & 7.5e-4 & 9.8e-3 & 8.6e-3 & 9.5e-3 & 5e-4 & --1.02 & 7.7e-4 & 2.6e-4 & 3.0e-3 \\
 15 & 8.4e-4 & 7.6e-4 & 1.5e-2 & 1.4e-2 & 9.4e-3 & 6e-4 & --0.95 & 7.7e-4 & 2.8e-4 & 4.7e-3 \\
\enddata
\tablecomments{
(1) Age of the SSP in Gyr.
(2) Iron mass locked in stars and remnants, assuming solar iron abundances 
    for the current stellar mass $M_*(t)$ in Table~\ref{tab2}.
(3) Remaining iron mass available to enrich the ICM.
(4) IMLR of the stellar component, computed from the
    luminosity $L_B(t)$ in Table~\ref{tab2}.
(5) IMLR of the ICM, provided all the remaining iron can escape the galaxy.
(6) Oxygen mass locked in stars and remnants, assuming [O/Fe]=+0.2 (see text).
(7) Remaining oxygen mass available to enrich the ICM.
(8) [O/Fe] ratio in the ICM.
(9) Silicon mass locked in stars and remnants, assuming [Si/Fe]=+0.2.
(10) Remaining Silicon mass available to enrich the ICM.
(11) SiMLR of the ICM, provided all the remaining silicon can escape the 
     galaxy.
 }
\end{deluxetable*}

In this paper, we shall consequently consider two cases for the stellar
metallicity, both with relative abundances 
[$\alpha$/Fe]=+0.2.
{\bf Case~A}: global metallicity, dominated by oxygen and $\alpha$--elements,
+0.2~dex, and consequently solar iron abundance;
{\bf Case~B}: global metallicity and $\alpha$--element abundances with solar 
values, and consequently depressed (below solar) iron abundance. 
Case~A seems more representative of the bulk of the stellar mass,
which resides in the central regions; Case~B minimizes the metals locked 
in the stars, within the range allowed by observations.

\begin{figure}
\epsscale{1.15}
\plotone{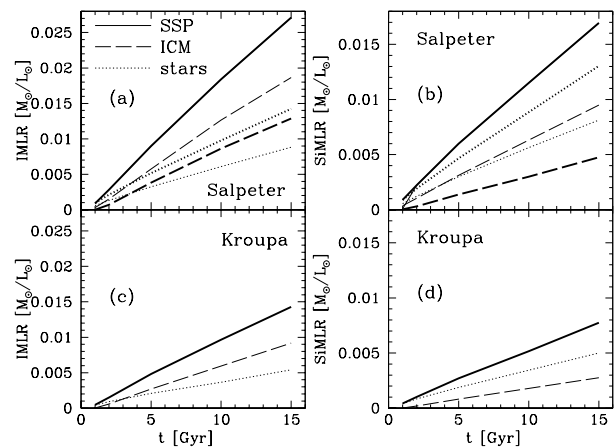}
\caption{Evolution of the IMLR and of the SiMLR for the global SSP ejecta
({\it solid lines}), for the ICM ({\it dashed lines}) and for the stellar 
component ({\it dotted lines}). Most of the evolution at advanced ages is due
to the fading luminosity of the SSP. The partition of metals between stars 
and ICM is computed \underline{self--consistently} from the M/L ratio of the 
SSP, assuming either solar iron abundance (Case~A, {\it thick lines}) or solar 
$\alpha$--elements abundance in the stars (Case~B, a ``minimal'' assumption,
{\it thin lines}). For the Kroupa IMF only Case~B is considered (panels c,d).
\label{fig:partIMLR-SiMLR} }
\end{figure}

In this Section we consider Case~A, corresponding to a solar
iron abundance in the stars $X_{Fe,*} = X_{Fe \odot} = 1.2 \times 10^{-3}$ 
(see footnote~\ref{foot:sol_ab} above). In Table~\ref{tab3} we list
as a function of age the resulting ${\rm MFe}_*$ and ${\rm MFe}_{ICM}$ 
derived from Eq.~\ref{eq:MFe*} and Eq.~\ref{eq:diffMFe} as a function of age.
From the typical luminosity $L_B(t)$ of the SSP (Table~\ref{tab2} and 
Fig.~\ref{fig:rateSN_MFe_LB_IMLRssp}c), we compute the corresponding IMLR 
for the stellar 
and ICM components respectively. The evolution of the global IMLR$_{SSP}$
of a Salpeter SSP, as well as of the separate components IMLR$_{ICM}$ and 
IMLR$_*$, is plotted in Fig.~\ref{fig:partIMLR-SiMLR}a (thick lines).

In standard Galactic Wind (GW) models for elliptical galaxies, the wind
occurs typically over timescales $\lesssim$1~Gyr, and after the GW
no further star formation or metal ejection from the galaxies occurs.
From Table~\ref{tab3}, if only the gas and metals produced 
within the first Gyr are ejected from the galaxy into the ICM, very little
iron (virtually none) is available to pollute the ICM, once the iron in the
stars is accounted for. For instance, the models of elliptical galaxies 
with Salpeter 
IMF in \cite{C2000, MPC03} are characterized by an early GW and eject 
a negligible
amount of iron in the ICM, so that the predicted IMLR$_{ICM}$ is an order 
of magnitude lower than observed.
Table~\ref{tab3} shows that, within the framework of the standard GW scenario,
the amount of iron that remains locked in the stellar component of a galaxy 
is not an effect of modelling details, but it is consistently determined 
by the IMF and the corresponding M/L ratio. Models with the Salpeter IMF 
and an early GW cannot release substantial amounts of iron to the ICM
--- if the observed solar stellar abundances are to be reproduced 
at the same time.

If, on the other hand, all the iron ever produced by the SSP over a Hubble 
time, and not locked in stars, is available to be dispersed into the ICM,
then it is possible for old SSPs ($t>10$~Gyr) to reach the observed levels of
IMLR$_{ICM}$ (Table~\ref{tab3} and Fig.~\ref{fig:obsICM}c, solid line).
This result corresponds to the original conclusion by \cite{MV88}
that, {\it if all the iron} produced by SN~Ia over long timescales escapes 
the galaxies, the observed iron mass and IMLR
in the ICM can be reproduced. \cite{MV88} predicted that the late SN~Ia ejecta 
are expelled into the ICM because, in the early galactic models,
the dominant dark matter contribution to the gravitational potential well 
was not included.
When the latter is accounted for, the consensus is that SN~Ia products 
cannot escape the galaxy \citep{DFJ91a, MG95}; although
recently \cite{Pip2002} could recover the original result of \cite{MV88}
by allowing for extreme efficiency in the energy feedback of SN~Ia.
Alternatively, the late SN~Ia ejecta could escape the galaxy via AGN feedback
or ram--pressure stripping \citep{R93}; see also \citet{Ci91} for the 
possibility of late outflows from elliptical galaxies.

\begin{deluxetable*}{cccccccccccc}
\tabletypesize{\scriptsize}
\tablecolumns{12}
\tablewidth{0pt}
\tablecaption{ {\mbox{Partition of metals between stars and ICM with the 
Salpeter IMF (Case B)}} \label{tab4}}
\tablehead{
\colhead{$^{(1)}$} & \colhead{$^{(2)}$} & \colhead{$^{(3)}$} & 
\colhead{$^{(4)}$} & \colhead{$^{(5)}$} & \colhead{$^{(6)}$} & 
\colhead{$^{(7)}$} & \colhead{$^{(8)}$} & \colhead{$^{(9)}$} &
\colhead{$^{(10)}$} & \colhead{$^{(11)}$} \\
\colhead{t[Gyr]} & \colhead{MO$_*$} & \colhead{MO$_{ICM}$} & 
\colhead{MFe$_*$} & \colhead{MFe$_{ICM}$} & \colhead{IMLR$_*$} & 
\colhead{IMLR$_{ICM}$} & \colhead{[O/Fe]$_{ICM}$} & \colhead{MSi$_*$} & 
\colhead{MSi$_{ICM}$} & \colhead{SiMLR$_{ICM}$}
}
\startdata
  1 & 6.5e-3 & 3.5e-3 & 5.8e-4 & 2.9e-4 & 5.9e-4 & 2.9e-4 &  +0.24 & 5.4e-4 & 3.5e-4 & 3.5e-4 \\
  2 & 6.2e-3 & 3.8e-3 & 5.6e-4 & 6.4e-4 & 1.3e-3 & 1.5e-3 & --0.07 & 5.2e-4 & 4.4e-4 & 1.0e-3 \\
  5 & 6.0e-3 & 4.0e-3 & 5.4e-4 & 9.6e-4 & 3.2e-3 & 5.7e-3 & --0.22 & 5.0e-4 & 5.2e-4 & 3.0e-3 \\
 10 & 5.9e-3 & 4.1e-3 & 5.3e-4 & 1.1e-3 & 6.1e-3 & 1.3e-2 & --0.27 & 4.9e-4 & 5.5e-4 & 6.3e-3 \\
 15 & 5.8e-3 & 4.2e-3 & 5.2e-4 & 1.1e-3 & 8.8e-3 & 1.9e-2 & --0.27 & 4.8e-4 & 5.6e-4 & 9.5e-3 \\
\enddata
\tablecomments{
(1) Age of the SSP in Gyr.
(2) Oxygen mass locked in stars and remnants, assuming solar oxygen (and 
    $\alpha$--elements) abundance for $M_*(t)$ in Table~\ref{tab2}.
(3) Remaining oxygen mass available to enrich the ICM.
(4) Iron mass locked in stars and remnants, assuming [O/Fe]=+0.2.
(5) Remaining iron mass available to enrich the ICM.
(6) IMLR of the stellar component, computed with the luminosity $L_B(t)$ 
    in Table~\ref{tab2}.
(7) IMLR of the ICM, provided all the remaining iron can escape the galaxy.
(8) [O/Fe] ratio in the ICM.
(9) Silicon mass locked in stars and remnants, assuming solar silicon (and 
    $\alpha$--elements) abundance for $M_*(t)$ in Table~\ref{tab2}.
(10) Remaining silicon mass available to enrich the ICM.
(11) SiMLR of the ICM, provided all the remaining silicon can escape the 
     galaxy.
 }
\end{deluxetable*}

\subsubsection{The role of the stellar M/L ratio}
\label{sect:MLratio}

Notice that, for the typical M/L ratio of an old Salpeter SSP and solar 
iron abundances in the stars (Case~A), an ``equipartition'' of iron between 
stars and ICM is predicted \citep{R93} 
--- a behaviour not mimicked, however, by $\alpha$--elements, 
see \S\ref{sect:abratios} and Fig.~\ref{fig:partOMLR} below.

It is the M/L ratio that determines, in the end, the amount of metals locked 
in the stars,
and consequently the partition of metals between galaxies and the ICM. In fact,
while stellar metallicities can be directly measured from colours and 
integrated spectral indices, the stellar mass can only be estimated
indirectly from
the observed luminosity and an assumed M/L ratio. If we are dealing with an 
old stellar population and a Salpeter IMF, the expected M/L ratio 
(Table~\ref{tab2}) is much larger than that estimated by \cite{W93} or by 
\cite{Bal01}, and often adopted to infer the partition in mass and metals 
between ICM and galaxies \citep{R97,R2003}. This is of minor relevance 
for the problem of the baryonic mass in clusters discussed by
\cite{W93}. From dynamical estimates, they adopt  
$M_*/L_B= 6.4 \, h$ and derive $M_*/M_{ICM} \sim$0.1 (for $h$=0.7); 
even with the M/L ratio expected from an old Salpeter IMF, 
a factor of $\sim$2 higher, the baryonic mass remains largely dominated 
by the hot ICM mass and the estimated baryon fraction in clusters is very 
little affected. However, the detailed assumption on the M/L ratio bears
much consequence on the estimated ``cold fraction'' $M_*/M_{ICM}$
\citep[a fundamental constraint on scenarios of cluster formation;][and 
references therein]{Torna03, MPC03}, as well as on the estimated
mass of metals locked in stars, and on the inferred effective yield in
clusters (see the Appendix).

With the low stellar M/L ratio from \cite{W93}, the metal 
partition is far more skewed toward the ICM, which would contain 2--3 times
more iron than galaxies \citep{R97,R2003}. However, if we assume so low
M/L ratios, we must conclude correspondingly that the IMF in clusters 
cannot be the Salpeter IMF but it is ``bottom--light'', meaning 
with a lower mass fraction locked--up 
in ever--lived low mass stars and remnants, and correspondingly with a 
higher net yield (see Eq.~\ref{eq:yield}).
It is thus decisive to discuss the chemical enrichment of the ICM with
yields and M/L ratios consistently derived from the same IMF (see also the
Appendix).

\begin{figure}
\epsscale{0.85}
\plotone{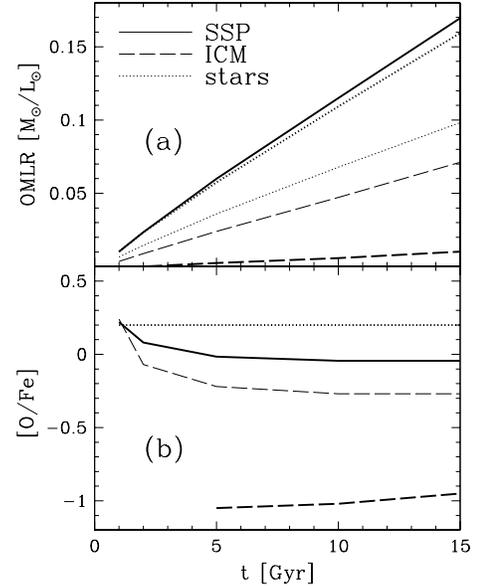}
\caption{(a) Partition of the Oxygen Mass to Light Ratio between stars and ICM,
for the Salpeter IMF; {\it thick lines}: Case~A for the stellar metallicities;
{\it thin lines}: Case~B for the stellar metallicities (the ``minimal'' 
assumption, see text). Notice that, while there can be 
equipartition of iron between stars and ICM (Fig.~\ref{fig:partIMLR-SiMLR}a) 
this is not the case for oxygen or $\alpha$--elements: with the Salpeter IMF
most of the oxygen produced is employed to build the observed stellar 
metallicities, and a very small fraction remains avaliable to enrich the ICM.
(b) Corresponding [O/Fe] ratio in the global ejecta, in the stars 
([O/Fe]=+0.2~dex, assumed) and in the ICM; with the
Salpeter IMF, the latter is strongly undersolar even in Case~B (the most 
favourable to the ICM enrichment).
\label{fig:partOMLR} }
\end{figure}

\subsubsection{The abundance ratios in the ICM}
\label{sect:abratios}

\noindent
We showed above that the Salpeter IMF can reproduce the 
observed iron enrichment if all the iron synthesized and not locked
in the stars is shed into the ICM. We will now discuss the enrichment
in $\alpha$--elements.

Under the assumption that star formation in ellipticals is a fast process with
a timescale $\lesssim$1~Gyr, as in standard GW models, the typical [O/Fe] ratio
of stars formed over such timescales is [O/Fe]=+0.2 (Table~\ref{tab1},
column~8 listing the typical [O/Fe] of the SSP ejecta up to an age $t$=1~Gyr;
see also Fig.~\ref{fig:partOMLR}b, solid line).
This is in good agreement with the latest observational estimates of 
the $\alpha$-enhancement in 
the bright elliptical galaxies which dominate the stellar mass in clusters
\citep{Trag2000a, Trag2000b, Mehl03}.
With a solar iron abundance in the stars (Case~A), [O/Fe]=+0.2 corresponds 
to an oxygen abundance
$X_{O,*}=1.33 \times 10^{-2}$ (see footnote~\ref{foot:sol_ab} above). 
With this value, from equations analogous to Eq.~\ref{eq:MFe*} 
and~\ref{eq:diffMFe} we calculate the oxygen mass MO$_*$ locked in stars, 
and a negligible amount MO$_{ICM}$ of the produced oxygen is left available 
to enrich the ICM (Table~\ref{tab3} and Fig.~\ref{fig:partOMLR}a, thick lines).
So, if the late SN~Ia
products are released into the ICM and the observed IMLR$_{ICM}$ is reproduced,
this inevitably leads to
strongly subsolar [$\alpha$/Fe] ratios in the ICM (Fig.~\ref{fig:partOMLR}b,
and Fig.~\ref{fig:obsICM}a for the case of silicon). 
This was in fact the 
original prediction by \cite{MV88}, as well as the recent result by 
\cite{Pip2002}. 
However, so low abundance ratios are at odds with observations:
low [$\alpha$/Fe] ratios are observed only in the very central regions of 
clusters, probing the ISM abundances in the central cD galaxy rather than
the ICM; the metallicity excess in ``cool cores'' (former cooling flows) 
is entirely due to the SN~Ia expected from the luminosity excess of the 
central bright galaxy and, in terms of mass, represent only some 10\%
of the total metal mass in the ICM \citep[][and references therein]{DeGra01,
DeGra03, Loe2003}.
At large radii, where the bulk of the ICM mass resides, abundance 
ratios are {\it at least} solar 
\citep[][and references therein]{F2000, Fin03, Baum03, R2003}.

We may alternatively assume Case~B for the stellar metallicity, with solar
values for the global metallicity and for {\Oxygen} and $\alpha$--element
abundances, and depressed iron abundance as following from the
typical [O/Fe]=+0.2. In this case less metals are locked in the stars and 
the enrichment of the ICM is favoured; 
the relevant stellar metallicities are 
$X_{O,*}=X_{O,\odot}=8.3 \times 10^{-3}$, $X_{Fe,*}=7.5 \times 10^{-4}$.
With these values in Eq.~\ref{eq:MFe*} and~\ref{eq:diffMFe},
we obtain the partition in Table~\ref{tab4} and Fig.~\ref{fig:partOMLR} 
(thin lines). 
The conclusions remain largely unchanged: the observed 
IMLR$_{ICM}$ can be reproduced only through late (after--GW) enrichment 
from SN~Ia; the consequent problem with the typical [O/Fe]$_{ICM}$ 
is somewhat relieved, yet the resulting [O/Fe]$_{ICM}$= --0.3 is too low 
with respect to observations (Fig.~\ref{fig:partOMLR}b, thin lines).

All in all, it seems very difficult to reproduce the observed IMLR$_{ICM}$
with a Salpeter IMF, without violating constraints on the abundance
ratios in the ICM. This is illustrated also in Fig.~\ref{fig:obsICM}a with the 
[Si/Fe] ratio (see \S\ref{sect:Silicon}): at the old ages where the 
IMLR$_{ICM}$ predicted with the Salpeter IMF reaches the observed levels
(Fig.~\ref{fig:obsICM}c), the corresponding
[Si/Fe] is negative. Our computations confirm the
well--known result that the Salpeter IMF predicts a ``chemical asymmetry'' 
with [$\alpha$/Fe]$>0$ in the stars and [$\alpha$/Fe]$<0$ in the ICM 
\citep{MV88, R93}. That no such asymmetry is
observed in real clusters, is very difficult to explain within a standard
chemical evolution scenario.

\subsection{The Silicon Mass--to--Light ratio}
\label{sect:Silicon}

\noindent
In the previous sections, we discussed the predictions of the Salpeter IMF
vs. observations for the case of iron and oxygen. Iron is the element
with the best measured abundance in the ICM; oxygen is the main
tracer of the global metallicity (roughly half of the global metals in the Sun)
as well as of the $\alpha$--elements, that is of the typical products of SN~II.
Observationally however, after iron the best measured element in the ICM 
is silicon \citep{Baum03}, also an $\alpha$--element mostly produced by 
SN~II --- 
although the contribution of SN~Ia in this case is not negligible. 
The observed Silicon Mass--to--Light Ratio (SiMLR) in the hot ICM is 
SiMLR$_{ICM} =$0.01--0.02~\Msol/\Lsol\ \citep{F2000, Fin03, Loe2003}.
In this Section we
compute and discuss the silicon production predicted from a Salpeter SSP.

With the chemical yields by \cite{PCB98},
the silicon production
from SN~II is somewhat metallicity dependent (with variations around 20\%), 
but a representative average value is 
\[ MSi_{II}=8 \times 10^{-4}~M_{\odot} \]
from a Salpeter SSP of 1~\Msol. This corresponds to 
[Si/Fe]=+0.5 as the typical
abundance ratio due to SN~II (see the adopted solar abundances in
footnote~\ref{foot:sol_ab} above). This is 
compatible with the plateau level in halo stars \citep{Cayr03}, 
or even +0.1~dex high in Si production.
Each SN~Ia produces 0.154~\Msol\ of silicon \citep{Iwa99}
so the SN~Ia in the SSP yield altogether
\[ MSi_{Ia} = 2.5 \times 10^{-4}~M_{\odot} \]
of silicon (Table~\ref{tab1}), 
a smaller but non--negligible contribution with respect to SN~II.
In the last three columns of Table~\ref{tab1} we list the silicon production 
$MSi_{Ia}$ from SN~Ia, total $MSi_{tot}$ from SN~Ia+SN~II 
and the [Si/Fe] ratio of the cumulated 
ejecta as a function of time. In the last column of Table~\ref{tab2} 
we compute the corresponding SiMLR$_{SSP}=MSi_{tot}/L_B$. In the last three 
columns of 
Table~\ref{tab3} we compute the partition of silicon between stars and the ICM,
with equations analogous to Eqs.~\ref{eq:MFe*} and~\ref{eq:diffMFe} and
assuming Case~A for the stellar metallicities.
The evolution of the global SiMLR$_{SSP}$ of a Salpeter SSP, as well as of the 
separate components SiMLR$_{ICM}$ and SiMLR$_*$, is plotted in 
Fig.~\ref{fig:partIMLR-SiMLR}b (thick lines).

From Fig.~\ref{fig:obsICM}b (solid line), the Salpeter SSP clearly falls short 
by at least a factor of 2 in explaining the observed SiMLR in the ICM.
Even minimizing the amount of metals locked in the stellar component by 
assuming Case~B 
(Table~\ref{tab4}), the predicted SiMLR$_{ICM}$ becomes at most
marginally consistent
with the lower limits of the the observed range, and only for very
large ages due to the very low luminosity (Fig.~\ref{fig:obsICM}b, 
dashed line). However
with the presently favoured cosmological scenario, Hubble constant value 
and corresponding age of the Universe, it seems unfeasible for the bulk
of the stellar population in clusters to be as old as $\sim$15~Gyr.

The too low resulting SiMLR is another signature that the Salpeter IMF
doe not produced enough $\alpha$--elements to enrich the ICM to the observed
levels.

\begin{deluxetable*}{cccccccccc}
\tablecolumns{10}
\tablewidth{0pt}
\tablecaption{ {\mbox{Metal production, luminosity and IMLR from 
a Kroupa SSP}} \label{tab5}}
\tablehead{ 
\colhead{$^{(1)}$} & \colhead{$^{(2)}$} & \colhead{$^{(3)}$} & 
\colhead{$^{(4)}$} & \colhead{$^{(5)}$} & \colhead{$^{(6)}$} & 
\colhead{$^{(7)}$} & \colhead{$^{(8)}$} & \colhead{$^{(9)}$} & 
\colhead{$^{(10)}$} \\
\colhead{t[Gyr]} & \colhead{$N_{Ia}$} & \colhead{$MFe_{tot}$} & 
\colhead{$MSi_{tot}$} & \colhead{${\cal M}_B$} & \colhead{$L_B$} & 
\colhead{$M_*$} & \colhead{$\frac{M_*}{L_B}$} & \colhead{IMLR$_{SSP}$} & 
\colhead{SiMLR$_{SSP}$} 
}
\startdata
  1 & 4.7e-4  & 6.6e-4 & 5.9e-4 & 5.135 & 1.385 & 0.82 & 0.59 & 4.8e-4 & 4.3e-4 \\
  2 & 9.2e-4  & 9.7e-4 & 6.6e-4 & 5.969 & 0.643 & 0.79 & 1.22 & 1.5e-3 & 1.0e-3 \\
  5 & 1.4e-3  & 1.3e-3 & 7.3e-4 & 6.912 & 0.270 & 0.75 & 2.76 & 4.8e-3 & 2.7e-3 \\
 10 & 1.5e-3  & 1.4e-3 & 7.5e-4 & 7.589 & 0.145 & 0.72 & 4.98 & 9.6e-3 & 5.2e-3 \\
 15 & 1.6e-3  & 1.4e-3 & 7.6e-4 & 8.008 & 0.098 & 0.71 & 7.20 & 1.5e-2 & 7.8e-3 \\
\enddata
\tablecomments{
(1) Age of the SSP in Gyr.
(2) Cumulative number of SN~Ia exploded up to age $t$.
(3) Total iron mass produced up to age $t$ by SN~II+SN~Ia.
(4) Total silicon mass produced up to age $t$ by SN~II+SN~Ia.
(5) B--band  magnitude of the SSP.
(6) B--band luminosity of the SSP.
(7) Mass fraction of the initial 1~\Msol\ SSP remaining in stars and remnants.
(8) M/L ratio of the stellar(+remnants) component of the SSP.
(9) Global IMLR of the SSP.
(10) Global SiMLR of the SSP.
}
\end{deluxetable*}

\begin{deluxetable*}{ccccccccc}
\tablecolumns{9}
\tablewidth{0pt}
\tablecaption{ {\mbox{Partition of metals between stars and ICM with the 
Kroupa IMF (Case B)}} \label{tab6}}
\tablehead{ 
\colhead{$^{(1)}$} & \colhead{$^{(2)}$} & \colhead{$^{(3)}$} &
\colhead{$^{(4)}$} & \colhead{$^{(5)}$} & \colhead{$^{(6)}$} &
\colhead{$^{(7)}$} & \colhead{$^{(8)}$} & \colhead{$^{(9)}$} \\
\colhead{t[Gyr]} & \colhead{MO$_*$} & \colhead{MFe$_*$} & 
\colhead{MFe$_{ICM}$} & \colhead{IMLR$_{ICM}$} & \colhead{MSi$_*$} & 
\colhead{MSi$_{ICM}$} & \colhead{SiMLR$_{ICM}$} & \colhead{[Si/Fe]$_{ICM}$}
}
\startdata
  1 & 6.8e-3 & 6.2e-4 & 4.4e-5 & 3.2e-5 & 5.7e-4 & 2.5e-5 & 1.8e-5 & --0.01 \\
  2 & 6.5e-3 & 5.9e-4 & 3.8e-4 & 5.9e-4 & 5.4e-4 & 1.2e-4 & 1.8e-4 & --0.27 \\
  5 & 6.2e-3 & 5.6e-4 & 7.3e-4 & 2.7e-3 & 5.1e-4 & 2.2e-4 & 8.0e-4 & --0.29 \\
 10 & 6.0e-3 & 5.3e-4 & 8.6e-4 & 5.9e-3 & 5.0e-4 & 2.6e-4 & 1.8e-3 & --0.28 \\
 15 & 5.9e-3 & 5.3e-4 & 9.0e-4 & 9.2e-3 & 4.9e-4 & 2.7e-4 & 2.8e-3 & --0.28 \\
\enddata
\tablecomments{
(1) Age of the SSP in Gyr.
(2) Oxygen mass locked in stars and remnants, assuming solar oxygen (and 
    $\alpha$--elements) abundance for $M_*(t)$ in Table~\ref{tab5}.
(3) Iron mass locked in stars and remnants, assuming [O/Fe]=+0.2.
(4) Remaining iron mass available to enrich the ICM.
(5) IMLR of the ICM, provided all the remaining iron can escape the galaxy.
(6) Silicon mass locked in stars and remnants, assuming solar silicon 
    (and $\alpha$--elements) abundance for $M_*(t)$ in Table~\ref{tab5}.
(7) Remaining silicon mass available to enrich the ICM.
(8) SiMLR of the ICM, provided all the remaining silicon can escape the galaxy.
(9) [Si/Fe] ratio of the ICM.
 }
\end{deluxetable*}

\subsection{A different star formation history?}

\noindent
Our computations are based on a simple SSP approach,
that can be a reasonable approximation for an early and rapid star formation 
history. In this case we showed that, if we require that the stars in 
elliptical galaxies have 
the observed solar metallicities and abundance ratios [$\alpha$/Fe]=+0.2,
it is difficult for the Salpeter IMF to reproduce the observed IMLR$_{ICM}$ 
without falling below solar [$\alpha$/Fe] ratios in the ICM 
(Fig.~\ref{fig:partOMLR}b and Fig.~\ref{fig:obsICM}).
One may wonder if agreement between the Salpeter IMF and observed abundance
ratios in the ICM is possible, once more complex star formation histories 
(SFH) are allowed for. In this Section we argue against this possibility.

Suppose to construct, with the Salpeter IMF, an {\it ad hoc} 
star formation history + wind/outflow history that produces a similar 
metal partition, and [$\alpha$/Fe] ratios, in the stars and in the ICM. 
This can only result in roughly solar ratios in both components, since 
the [$\alpha$/Fe] ratio of the global metal production is solar
(Table~\ref{tab1} and Fig.~\ref{fig:partOMLR}b). Though it is doubtful that 
solar abundance ratios are representative of the bulk of stars in cluster 
ellipticals, we shall assume it for the sake of argument.

The required picture is likely to be a prolongued star formation history 
with a concurrent continuous partial outflow of metals. Stars 
keep forming locking up a sizeable fraction of the delayed SN~Ia products, 
while at the same time part of the metals keep escaping into the ICM. 
Such a prolonged SFH 
contrasts with the evidence that the bulk of the stellar population 
in ellipticals is old. But more important, the presence of a younger stellar 
component
would substantially reduce the global IMLR$_{SSP}$ due to the increased 
luminosity (Table \ref{tab3} and Fig.~\ref{fig:rateSN_MFe_LB_IMLRssp}). W
e conclude it would be unfeasible, with the Salpeter IMF,
to reconcile a balanced distribution of metals and abundance ratios between
stars and ICM, maintaining a high global IMLR$_{SSP}$ at the same time.

\section{Predictions with a ``standard'' IMF}
\label{sect:Kroupa}

\noindent
All the calculations in the previous Section~\ref{sect:Salpeter} were 
performed for a Salpeter
IMF with mass limits [0.1--100]~\Msol. This is often erroneously considered 
as a ``standard'' IMF (in the sense explained in the introduction),
but as anticipated in Section~\ref{sect:Salpeter} it is in fact {\it already 
too efficient} in metal enrichment, i.e.\
it has too high typical chemical yields (especially for oxygen and
$\alpha$--elements), 
to match the typical abundances in the Solar Neighbourhood within the standard 
chemical evolution scenario
(including infall and no outflows). With a Salpeter slope, an upper mass limit
of 50--70~\Msol, corresponding to a smaller number of SN~II and lower metal
production, should rather be adopted to reproduce observations of the 
Solar Neighbourhood and spiral galaxies in general
\citep{Tsu97,TGB98,Gra2000,PSLT03}. The \cite{Sca86} or \cite{Kr98}
IMF, with a steeper
slope at the high-mass end and lower yields, are better suited to model
the local environment \citep{CMG97,PCB98,BP99}.

The difference in (oxygen) yield between a Salpeter and a Kroupa IMF 
with the same mass limits [0.1-100]~\Msol\ 
is a factor of $\sim$1.7 \citep[cf.\ Table~3 of][]{PSLT03}. 
This difference 
is readily explained by considering Eq.~\ref{eq:yield}: the shape of the IMF
influences the net yield via the ratio $\zeta_9/(1-R)$, where $\zeta_9$
is the amount of mass in stars with $M>9$~\Msol, which become SN~II and 
produce the bulk of the metals;
see \citet{PSLT03} for a detailed discussion. The Salpeter and Kroupa IMFs
have comparable locked--up fractions $1-R \sim$70\%, but due to the steeper 
slope at the high--mass end, $\zeta_9$ is about twice lower for Kroupa than
for Salpeter \citep[cf.\ Table~3 of][]{PSLT03}.
Henceforth, even if
the Salpeter IMF were considered to be marginally consistent with the observed
enrichment in clusters (disregarding the evidence from $\alpha$--element
abundances), the implied metal production is already twice higher
than that of standard IMFs suited to reproduce the Milky Way.

From Fig.~\ref{fig:rateSN_MFe_LB_IMLRssp} we can immediately see that the 
global iron production of a Kroupa SSP is also lower than that of a Salpeter
SSP. At the same time, the typical luminosity of a Kroupa SSP is larger,
because it has less mass locked in very low mass stars and an increased 
percentage of ``intermediate--living stars'' (around 1~\Msol) with respect 
to the Salpeter IMF; the latter mass range is generally responsible
for the bulk of the luminosity \citep{PSLT03}.
As a consequence of the lower iron yield and higher luminosity,
the resulting Kroupa IMLR$_{SSP}$ is substantially lower than for the 
Salpeter SSP. Therefore, we can anticipate that the Kroupa IMF will hardly
be able to produce the observed IMLR in clusters of galaxies.

\begin{figure}
\epsscale{1.08}
\plotone{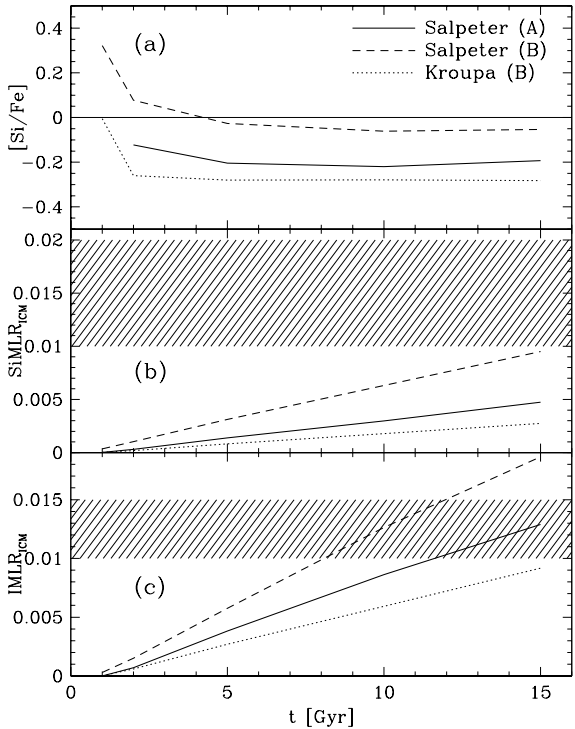}
\caption{Predicted IMLR, SiMLR and [Si/Fe] ratio in the ICM, as a function
of the age of the stellar population; the shaded areas represent the 
observational range. All the metals not locked in the stars 
are assumed to be able to escape the galaxies and enrich the ICM. {\it Solid
line}: Salpeter IMF with Case~A for the stellar metallicities; 
{\it dashed line}: Salpeter IMF with Case~B for the stellar metallicities
(the ``minimal'' assumption, see text); 
{\it dotted line}: Kroupa IMF with Case~B. The Salpeter IMF can reach
the observed level of IMLR at old ages (c), however the predicted [$\alpha$/Fe]
ratios are then significantly undersolar (a), at odds with observations.
This reflects in the low predicted SiMLR, steadily below the observational 
level (b).
With the ``standard'' Kroupa IMF, the predicted ICM enrichment is
way below the observed levels at all times for both iron and silicon (b,c). 
\label{fig:obsICM} }
\end{figure}

In this Section we compute the expected chemical enrichment of the ICM
from a really standard IMF \citep{Kr98} with the same procedure we applied
to the Salpeter IMF.
Table~\ref{tab5} and Table~\ref{tab6} are analogous
to Table~\ref{tab1} to~\ref{tab4} but for an SSP with the Kroupa IMF, 
and the results clearly show that with 
this really ``standard'' IMF one cannot explain the observed IMLR or SiMLR 
in clusters of galaxies (Fig.~\ref{fig:obsICM}, dotted line). 
With the yields by \cite{PCB98}, a 1~\Msol\ Kroupa SSP produces 
\[ MO_{II}=5.7 \times 10^{-3}~M_{\odot} \]
\[ MFe_{II}=3.3 \times 10^{-4}~M_{\odot} \] 
\[ MSi_{II}=5.2 \times 10^{-4}~M_{\odot} \] 
of \Oxygen, \Fe\ and \Silicon\ from SN~II. 
In Table~\ref{tab5}
we compute the rate of SN~Ia, the global metal production and the luminosity 
evolution
for a Kroupa SSP. 
It is evident from Fig.~\ref{fig:rateSN_MFe_LB_IMLRssp}d and
the tabulated IMLR$_{SSP}(t)$ in Table~\ref{tab5}, 
that the Kroupa IMF may match the observed IMLR$_{ICM} \sim$0.015
only for very high ages and provided {\it all} the iron produced
is used to enrich the ICM with none left in the stellar component, which is
unreasonable. The SiMLR$_{SSP}$ is at all times lower than the levels $>$0.01
observed in the ICM, which can never be matched 
(Fig.~\ref{fig:partIMLR-SiMLR}d).

In Table~\ref{tab6} we compute the partition of metals between stars and ICM.
It is sufficient to consider only Case~B for the stellar metallicities (solar 
$\alpha$--element abundances and [$\alpha$/Fe]=+0.2) since this minimizes 
the metals locked in the stars and allows for a more efficient enrichment 
of the ICM. The resulting IMLR and SiMLR for the global SSP evolution, and
the partition between stars and ICM, are plotted in 
Fig.~\ref{fig:partIMLR-SiMLR}cd.

Even with the most favourable Case~B assumption, 
the Kroupa IMF barely produces the oxygen mass needed to enrich the stars
(column~2 in Table~\ref{tab6}),
hence virtually no oxygen is left available to enrich the ICM.
As anticipated above, the predicted IMLR$_{ICM}$ becomes
only marginally consistent with the observed one for extremely large ages,
and then the corresponding [$\alpha$/Fe] ratio in the ICM is well subsolar,
at odds with observations. The latter feature is reflected in the very
low SiMLR$_{ICM}$, lying below the observed one by a factor of 3 or more
(Fig.~\ref{fig:obsICM}, dotted lines).

\section{Discussion}
\label{sect:discussion}

\noindent
In this paper we computed the metal production expected with a Salpeter IMF 
and the corresponding evolution of the global IMLR and SiMLR 
(Fig.~\ref{fig:partIMLR-SiMLR}ab, solid line). Then we computed the 
corresponding partition of metals and IMLR between stars and ICM 
(Fig.~\ref{fig:partIMLR-SiMLR}ab, dotted and dashed lines).

We demonstrated that it may be possible to reproduce the observed IMLR 
in the ICM (Fig.~\ref{fig:obsICM}c, shaded area) with a Salpeter IMF, provided 
all the metals ever produced
and not locked in the stars are expelled from the 
galaxy and contribute to enrich the ICM, including the late SN~Ia contribution.
However, the SiMLR is not reproduced and the predicted [$\alpha$/Fe] ratios 
are significantly 
undersolar, as shown in Fig.~\ref{fig:obsICM}ab
\citep[cf.\ the ``chemical asymmetry'',][]{R93}. This is at odds with 
observations.
These results obtained with a straightforward, simple computation 
qualitatively agree with results from more elaborated models following the
detailed star formation and wind history of elliptical galaxies
\citep{MV88,Pip2002}.

We remark also that the Salpeter IMF is {\it not} a ``standard'' IMF 
in the sense that it is not well suited to model the chemical evolution
of the Solar Neighbourhood; a Scalo or Kroupa IMF
provides better predictions. 
On the other hand, with these IMFs it is impossible to account 
for the level of metal enrichment observed in clusters of galaxies,
and not only for the $\alpha$--elements but also for iron
(Fig.~\ref{fig:obsICM}, dotted lines).
We also stress that it is not just the (possibly still uncertain) 
{\it abundance ratios}, but the {\it global} amount of metals observed 
in the ICM that points to a different IMF with respect to the Solar 
Neighbourhood.

\subsection{Role of uncertain stellar yields and supernova rates}
\label{sect:uncertainties}

\noindent
The results discussed above are robust with respect to uncertainties 
on (a) stellar and SN yields, and (b) the rate of SN~Ia.

\begin{description}
\item[(a)]
One may argue that if theoretical nucleosynthesis calculations (the stellar 
yields $p_Z(M)$ in Eq.~\ref{eq:yield}) are uncertain by an overall factor of 
$\sim$3, we may be grossly underestimating the metal production 
of any given IMF. Then, even a local (Scalo or Kroupa) IMF could produce the 
observed enrichment in galaxy clusters --- which is a factor of 3 higher 
than that estimated in the Solar Neighbourhood 
(Eq.~\ref{eq:yield_SV}--\ref{eq:yield_cl} and Appendix).

Theoretical uncertainties in stellar yields of SN~II may be important to 
establish the {\it relative} contribution of SN~II and SN~Ia in the enrichment
of the ICM \citep{GLM97} --- though the role of this uncertainty 
may be reduced, when one considers only those theoretical SN~II yields 
that reproduce the 
observed tight [$\alpha$/Fe] plateau in halo stars \citep{Cayr03}.
In terms of
{\it global} metal production, however, the uncertainty in SN~II yields
seems to be below the
required factor of 3 (see the Appendix); at least if we can estimate the real 
uncertainty in SN~II models from comparing different sets of available 
SN yields.

More important, if stellar yields were a factor of 3 higher than adopted here, 
the same Scalo or Kroupa IMF could be hardly reconciled with the observed
chemical evolution in the Solar Neighbourhood. We underline in fact that
the stellar yields adopted here do reproduce the Solar Neighbourhood
when combined with a Scalo/Kroupa IMF \citep{PCB98}, and that we scaled 
the most uncertain iron yield from SN~II so as to reproduce properly the 
observed [$\alpha$/Fe] plateau in halo stars (\S\ref{sect:metals_salp_ssp}
and~\S\ref{sect:Silicon}). This makes our comparison between Solar 
Neighbourhood
and cluster enrichment as independent as possible of formal uncertainties 
in SN~II yields, since we are using stellar yields that have been tested, 
and partly calibrated, to reproduce the Solar Neighbourhood with the local 
Scalo/Kroupa IMF. 

Besides, it is comforting that a number of different groups have modelled 
the Solar
Neighbourhood independently, each with their favourite set (or patchwork
of sets) of stellar yields, and there is a general consensus that a 
Kroupa/Scalo IMF well reproduces the properties of the Solar Vicinity
\citep{MF89,CMG97,PCB98,BP99,Ali2001,Liang01,deDon2002,deDon2003}, while 
a Salpeter IMF gives worse results both 
for the Solar Neighbourhood \citep{Tsu97,TGB98,Gra2000} and for disc galaxies 
in general \citep{PSLT03}.
It seems that the gross overall picture for the Solar Neighbourhood
within standard chemical evolution models is well settled, in spite of the 
potential theoretical uncertainties in SN~II yields. 

If real stellar yields were significantly higher than the commonly adopted 
nucleosynthesis 
prescriptions in literature, to explain the Solar Neighbourhood with such 
an excess of metal production one should invoke substantial metal losses 
from the disc --- a non--standard ingredient for
chemical models of the Milky Way (see \S\ref{sect:conclusions}).

\item[(b)] 
Uncertainties about the progenitors of SN~Ia
and in their evolution may well allow for SN~Ia rates different from the
standard \cite{GR83} formulation, adopted here and in most literature. 
With a different formulation, possibly the SN~Ia rate
in ellipticals was higher than predicted here, and/or more SN~Ia ejecta
could enrich the ICM to high levels of {\it iron}, even with a standard IMF. 
However, any attempt to compensate for the predicted low enrichment just 
by modifying the SN~Ia rate, would not contribute significant 
$\alpha$--elements and would decrease the [$\alpha$/Fe] ratio in the 
ICM further below solar values, at odds with observations. Therefore,
adjusting the SN~Ia contribution is not a viable solution to reconcile
a standard IMF with the metal content in clusters.
\end{description}

\subsection{How is the IMF in clusters ?}
\label{sect:clusterIMF}

\noindent
It is evident that the IMF in clusters of galaxies cannot be the Salpeter IMF
--- and even less so a standard Solar Neighbourhood IMF. In this Section we
try to outline the qualitative features necessary for an IMF able to produce
the observed chemical enrichment in clusters. 

The net yield of a stellar population (Eq.~\ref{eq:yield}) depends on the 
shape of the IMF essentially via the ratio $\zeta_9/(1-R)$, between the mass 
fraction $\zeta_9$ ``invested'' in the progenitors of core--collapse 
supernov\ae\ that contribute the bulk of the metals ($M\geq9$~\Msol),
and the locked--up fraction $(1-R)$
\citep[see][for a detailed discussion]{PSLT03}. The net yield can therefore
be increased by increasing the number of massive stars and/or by reducing
the locked--up fraction.

With respect to Salpeter, the cluster IMF should lock less mass in very 
low--mass stars, with a correspondingly lower locked--up fraction and lower 
M/L ratio. The latter feature would likely be in better agreement also with 
determinations of the dynamical M/L ratio in ellipticals \citep{Ger01, Borr03}.
With such a ``bottom--light'' IMF the typical IMLR$_{SSP}$, the
relative numbers of SN~II and SN~Ia and the corresponding global 
[$\alpha$/Fe] ratio of the stellar ejecta, are not necessarily different from
the Salpeter IMF. What changes is the global yield (Eq.~\ref{eq:yield}) and 
the fraction of 
metals locked in stars, hence the partition of metals between the galaxies 
and the ICM is skewed in favour of the latter. In this case, within the early 
GW scenario enough metals may be dispersed into the ICM reproducing the 
observed IMLR$_{ICM}$, while preserving solar 
or moderately supersolar [$\alpha$/Fe] both in the stars and in the ICM.

However, in clusters the global [$\alpha$/Fe] ratio in the baryons seems to be
somewhat supersolar --- supersolar in the elliptical galaxies, at least solar
in the ICM --- therefore the relative SN~II/SN~Ia proportion is likely to be
somewhat higher than that of the Salpeter IMF, which has typical solar ratios
in the ejecta (\S\ref{sect:metals_salp_ssp}). Somewhat shallower slopes than
the Salpeter for $M>$1~\Msol\ are favoured in this sense, although the 
quantitative details may depend on the assumed SN~II yields \citep{GLM97}.

Definitely, the high--mass slope needs to be shallower than the Kroupa/Scalo 
one: the Kroupa IMF, in fact, in spite of being ``bottom--light'' with respect 
to Salpeter below 1~\Msol, does not produce enough metals to explain clusters.
Therefore, a larger $\zeta_9$ (larger nuber of massive stars) is needed for
the cluster IMF, than in the Kroupa IMF.

These qualitative features --- high overall $\zeta_9/(1-R)$ --- are 
achieved, for instance, with the variable IMF scenario considered by 
\citet{C2000, MPC03}
where the low--mass cut-off is skewed toward higher masses, and the high-mass
slope tends to be shallower, at high redshifts and in larger galaxies; this 
scenario has also a number of advantages in explaining the spectro--photometric
properties of ellipticals \citep{C98}.
Possibly, a similar result for the enrichment of the ICM can be obtained with 
an invariant IMF, provided it is characterized by a sufficiently high net
yield (i.e.\ high $\zeta_9/1-R$) as described above. 

\section{Conclusions : two alternative scenarios}
\label{sect:conclusions}

\noindent
From the previous discussion it is clear that a standard Solar Neighbourhood
IMF cannot explain the chemical enrichment of clusters of galaxies. 
We are then left with only two possible, alternative conclusions.
\begin{enumerate} 
\item
The IMF is different locally than in clusters. The variation probably
depends on redshift as well as on local density and temperature conditions,
as expected from theoretical arguments \citep[][and references therein]{MPC03}.
Such a scenario could
also explain the systematic differences between rich clusters and groups
\citep{Fin03}, and the increasing level of $\alpha$--element ICM enrichment 
with cluster temperature \citep{Fuka98, Fuka2K, F2000, Baum03}.
\item
The IMF is universal and invariant, but from cluster observations it must be
far more efficient in metal production than we think and commonly assume 
in chemical models for the Milky Way. If this is the case, since we do not 
observe such high metallicities in disc galaxies, we must invoke
substantial outflows of metals from spirals, too --- and not only from the
old bulge component resembling elliptical galaxies \citep{R2002,R2003}, 
but also throughout the disc evolution.
\end{enumerate}
It seems the choice is between a non--standard IMF or a non--standard
scenario for the chemical evolution of the Solar Neighbourhood and discs
galaxies. We will discuss the feasibility
of either possibility in turn, here below.

\subsection{A non--standard IMF in clusters}

\noindent
As discussed in \S\ref{sect:clusterIMF}, the key characteristic of the cluster
IMF should be a low locked--up fraction, i.e.\ a smaller number of ever-lived,
low--luminosity stars than in the Salpeter IMF; combined with a larger number 
of massive stars and
SN~II than expected from the Scalo/Kroupa slope. We outline here some 
possibile physical mechanisms to obtain a non--standard IMF in clusters, 
with these characteristics.

\begin{description}
\item[(i)]
At high redshifts, the increasing temperature of the cosmic micro-wave 
background and the lower gas metallicity (i.e.\ reduced cooling efficiency) 
contribute to increase the temperature of the star--forming gas. Higher
temperatures imply lower Mach numbers and less efficient turbulent compression,
which is crucial to trigger the formation of low--mass stars.
At increasing redshift, the Jeans mass of the compressed gas clumps is expected
to increase and the formation of low--mass stars is consequently hampered
\citep{NP2003}.
In clusters, the bulk of star formation takes place at high redshift in 
elliptical galaxies, 
which favours a lack of low--mass stars with respect to Solar Neighbourhood
conditions.

A detailed discussion of the chemical evolution of the ICM within the 
scenario of a variable Jeans mass can be found in 
\citet{MPC03}. A redshift dependence of the Jean mass could also explain 
the systematic differences between rich clusters and groups \citep{Fin03}.
However, a more drastic variation than that expected from the sole
cosmic background temperature seems to be required, to reproduce the
metal enrichment in clusters \citep{MPC03,Fin03}.

It is interesting that in starburst galaxies, with huge star formation rates 
resembling those that must have been typical of the early elliptical galaxies,
dynamical arguments favour low--mass cut-offs around 1--5~\Msol, 
i.e.\ the absence of
significant mass in low--mass stars \citep{Lei98}.

\item[(ii)]
The mass of the largest star formed in a star cluster increases with total 
cluster mass, just as expected if stellar masses are drawn
from a declining power--law probability distribution. The masses of star
clusters follow themselves a mass function roughly $\propto M^{-2}$. The
convolution of these two effects can explain the observed difference
in the high--mass IMF slope between star clusters (Salpeter slope)
and field (Scalo slope); see \citet{KrWei03, Kr2003}.

Large star clusters are more likely to form in large galaxies and/or in
regimes of high star formation rate; in such system, the net effect is
to produce a larger number of massive stars, although the 
IMF distribution within each star cluster remains invariant
\citep{KrWei03}. This effect can induce a more top--heavy {\it effective} IMF
in ellipticals, where star formation is rapid and intense, with respect to 
discs, where the star formation history is smoother; 
hence an instrinsic difference is expected between clusters and
disc galaxies like the Milky Way. Also, the IMF would be more top--heavy
in massive galaxies with respect to smaller galaxies; this latter trend is 
the same proposed, via a variable IMF, by \citet{C98} to explain the ensemble 
of photometric properties of elliptical galaxies.
\end{description}

\subsection{A universal IMF and metal outflows from the Disc}
\label{sect:concl_outflows}

\noindent
If, alternatively, the IMF is universal and invariant, then to explain
the observed metal enrichment in clusters its typical metal production 
(global yield) must be more efficient than usually assumed in the Solar 
Neighbourhood, by a factor of 3 (Eq.~\ref{eq:yield_SV}--\ref{eq:yield_cl} and
Appendix).

In \S\ref{sect:clusterIMF} we argued for a ``bottom--light'' IMF
in clusters (with respect to Salpeter). A bottom--light IMF is in fact
favoured also in the Solar Neighbourhood from local star counts 
\citep{Kr2001,Kr2002,Cha2003} and in disc galaxies in general, from
arguments related to their low stellar M/L ratio and the observed 
brightness of the Tully--Fisher relation \citep{PSLT03}. 
The evidence that the IMF is everywhere ``bottom--light'' 
with respect to Salpeter below 1~\Msol\ also reduces, by about
a factor of 2, the inferred star formation rate from the observed luminosities
\citep{Lei98,Cole01}.

The global yield of 
bottom--light IMFs tends to be high, due to the little mass locked in low--mass
stars (Eq.~1). At the same time, the IMF slope in the range 
of massive stars needs to be shallower than the Kroupa/Scalo slopes: these 
standard IMFs in fact, in spite of being bottom-light with respect to Salpeter,
have been shown in this paper to be unable to produce enough metals to explain
clusters.
The IMF slope in the range of massive stars ($M>$10~\Msol) determined in
nearby young clusters and associations as well as in starburst galaxies is
generally consistent with a Salpeter slope, therefore shallower
than the Scalo one \citep{Lei98,Mass98}.
Also the observed cosmic rate of SN~II favours shallower slopes than the
Scalo/Kroupa one, for a universal IMF \citep{Cal2003}.

Notice that shallower slopes than the Scalo one are compatible with
observational determinations of the local IMF, since the high---mass slope 
is poorly determined 
and strongly depends on corrections for the binary fraction; slopes up to 
--1.3, about the Salpeter value, are possible \citep{Kr2001,Kr2002}.
Besides, global properties of spiral galaxies (notably, the H$_\alpha$ 
emission) seem to be better explained with shallower slopes than the Scalo 
one \citep{KTC94, SL96}.
It is in fact intriguing that 
some of the bottom--light IMFs considered by \cite{PSLT03} to model the stellar
M/L ratio in disc galaxies, do have yields comparable to the observed yield 
in clusters.

A higher global yield in the Solar Neighbourhood can also be obtained
maintaining a standard Kroupa/Scalo IMF, but invoking a significant 
(a factor of 3) underestimate in the stellar yields of massive stars;
though we do not particularly favour this possibility 
(see \S\ref{sect:uncertainties} and the Appendix).

Irrespectively of the favoured way (IMF or stellar yields) to obtain locally 
a global yield as high as in clusters, we do not observe that much metals
in the Solar Neighbourhood (cf.\ Eq.~\ref{eq:yield_SV} 
and~\ref{eq:yield_cl}) and a universal IMF that explains clusters necessarily 
requires substantial outflows of metals from galactic discs: at least 70\%
of the metals produced should be ejected (see the Appendix).

We foresee the following difficulties with this scenario.
In discs, star formation proceeds at a smooth, non burst--like pace
and although ``fountains'' and ``chimneys'' are observed, they do not have 
the necessary energy to escape the galactic potential
\citep{Breg80, DS86, 
Heck2002}; therefore winds are far less plausible in galactic discs than 
in spheroids. Although recent results indicate that winds are
more common and ubiquitous than previously thought, even in our Galaxy and
in normal star--forming galaxies \citep{Veill03}, what one observes
is usually hot gas that escaped the disc; it is by no means obvious
that this gas can escape into the inter--galactic medium altogether,
and not just fall back onto the galaxy at some later time.

Moreover, from the dynamical point of view, strong ongoing 
stellar feedback and outflows could significantly hamper the progressive
formation of galactic discs from the cool--out of halo gas 
\citep[][Sommer--Larsen, Portinari \& Laursen in prep.]{SLGP03}

Finally, even though we may regard substantial metal outflows
as a viable scenario, they certainly represent a ``non--standard picture'' 
for the chemical evolution of the Solar Neighbourhood, which would demand
a revision of our understanding of the formation and evolution 
of the disc of the Milky Way.

\acknowledgments
\noindent
We benefitted from discussions with K.~Pedersen, J.~Andersen, {\AA}.~Nordlund, F. Calura and M. Riello. LP acknowledges kind hospitality from the Astronomy Department
in Padova and from the Observatories of Helsinki and Tuorla upon various 
visits.
This study is financed by the Danmarks Grundforskningsfond (through 
the establishment of TAC) and by the Italian MIUR.

\appendix
\section{Observed effective yields in the Solar Neighbourhood and in clusters}

\noindent
In this Appendix we will better quantify the difference
between the effective yields observed in clusters and in the Solar 
Neighbourhood, respectively (cf.\ Eq.~\ref{eq:yield_SV} and~\ref{eq:yield_cl}).
In the hypothesis that the effective yield in clusters is representative of a
universal IMF, this estimate will give us an idea of how much metals must have
outflown from the Solar Neighbourhood to the intergalactic medium, to reconcile
with the low metal content observed locally.
We will discuss iron and $\alpha$--elements separately.
\subsection{The Solar Neighbourhood}

\noindent
The local surface mass density is about 50~\Msol~pc$^{-2}$, with negligible
dark matter contribution \citep{KG91,FF94,Sack97}. The local gas surface 
density is about 10~\Msol~pc$^{-2}$ \citep{Dame93}. Henceforth, locally
$M_{gas}=20-25\% \, M_*$.

The iron abundance distribution of local stars peaks between --0.2 and 
--0.1~dex
\citep{WG95, Rocha96, Hou98, JorBR2000, FM97, Kot02}. Assuming 
--0.15~dex (or $Z_{Fe,*} = 0.7 \, Z_{Fe, \odot}$) as
a representative value for the stellar iron abundance, and assuming a solar 
abundance in the gas, the observed iron yield in the Solar Vicinity is:
\begin{equation}
\label{eq:yieldFe_SV}
y_{Fe, SV} \sim \frac{0.7 \, Z_{Fe, \odot} \times M_* + Z_{Fe, \odot} \times 
(0.2-0.25 \, M_*)}{M_*}
 = 0.9 - 0.95 \, Z_{Fe, \odot}
\end{equation}
As to the $\alpha$--elements, the bulk of local disc stars range between 
[O/Fe]=--0.1 to +0.3~dex, and between [Si/Fe]=0 to +0.2~dex \citep{Edv93}.
We can take therefore [$\alpha$/Fe]=+0.1~dex as a representative average value;
also, for [Fe/H]=--0.15~dex where the metallicity distribution peaks, 
[$\alpha$/Fe]$\sim$+0.1 in the data by \citet{Edv93}. This implies
{\mbox{[$\alpha$/H]=--0.05~dex}}, or $Z_{\alpha,*} =0.9 \, Z_{\alpha, \odot}$ 
as a typical value for 
disc stars. Assuming still solar abundances for the gas, the local effective
yield in $\alpha$--elements is:
\begin{equation}
\label{eq:yieldalfa_SV}
y_{\alpha, SV} \sim \frac{0.9 \, Z_{\alpha, \odot} \times M_* + 
Z_{\alpha, \odot} \times (0.2-0.25 \, M_*)}{M_*}
 = 1.1-1.15 \, Z_{\alpha, \odot}
\end{equation}
Therefore, the observed effective yield in the Solar Vicinity is about
solar (within 10\%) both for iron and for $\alpha$--elements.

Notice that locally, the baryonic mass is dominated by the stellar mass,
hence the effective yield is very close to the stellar metallicity, while
the metals in the gaseous phase give a small contribution.
Henceforth, the
uncertainty in the gas metallicity and in the $M_{gas}/M_*$ ratio does not
affect much the local effective yield. The situation is very different
in clusters, where the gas mass dominates and the estimated $M_{gas}/M_*$ 
ratio is crucial to determine the effective yield.
\subsection{Clusters of galaxies}

\noindent
From their accurate BeppoSAX iron abundance profiles, 
deprojected and convolved with the gas density profiles, \citet{DeGra03}
derive an average iron abundance by mass in the ICM 
$Z_{Fe, ICM} = 0.34 \, Z_{Fe,\odot}$ \citep[0.23~$Z_{Fe, \odot}$ in the old 
photospheric scale by][]{AG89}.
The typical iron abundance in the stellar populations in clusters is between
[Fe/H]=--0.2 to 0~dex, or $Z_{Fe, *} = 0.6-1 \,Z_{Fe, \odot}$ (Case B and A 
considered in this paper, respectively; see \S\ref{sect:partition}).

\begin{figure}
\epsscale{0.52}
\plotone{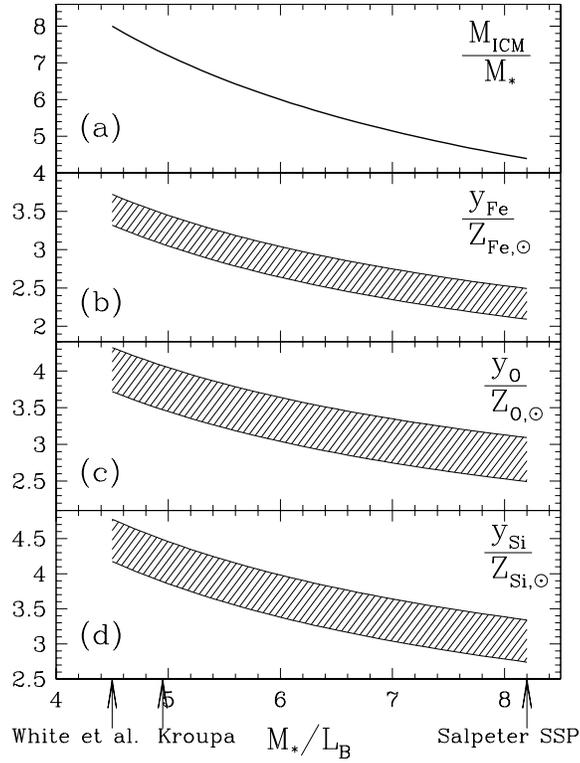}
\caption{(a) ICM--to--star mass ratio, estimated from the observed
 $M_{ICM}/L_B = 36$~\Msol/\Lsol\ as a function of the assumed stellar M/L 
ratio.
(b) Corresponding iron effective yield, in solar units; the vertical range
corresponds to different assumptions for the stellar metallicities (see text).
(c) Corresponding oxygen effective yield, in solar units.
(d) Corresponding silicon effective yield, in solar units.
\label{fig:yield_cl} }
\end{figure}

The crucial and most uncertain quantity entering the estimate of the effective
yield is the mass in the stars, since in clusters this can only be estimated
indirectly from the observed luminosity --- hence the importance of accounting
self--consistently for the stellar M/L ratio, as we underline in this paper.
\citet{MPC03} noticed that different samples in literature agree very well
on a typical value $M_{ICM}/L_B = 30 \, h^{-\frac{1}{2}} = 36$~\Msol/\Lsol,
while the derived $M_{ICM}/M_*$ can vary a lot after the assumed M/L ratio.
In Fig.~\ref{fig:yield_cl}ab
we show, as a function of the 
stellar M/L ratio, the corresponding $M_{ICM}/M_*$ ratio and the effective 
iron yield, calculated as:
\begin{equation}
\label{eq:yieldFe_cl}
y_{Fe, cl} \sim \frac{(0.6-1 \, Z_{Fe, \odot}) \times M_* + 
0.34 \, Z_{Fe, \odot} \times M_{ICM}}{M_*}
\end{equation}
The stellar M/L ratios in Fig.~\ref{fig:yield_cl} range between the
typical value for a Salpeter SSP which is 10~Gyr old (a minimal age to reach
the observed IMLR, \S\ref{sect:Salpeter}) and the extensively quoted
value 6.4~$h$=4.5~\Msol/\Lsol\ estimated by \citet{W93}. The Salpeter value 
should be considered
as an upper limit to the M/L ratio (and consequently a lower limit to
the $M_{ICM}/M_*$ ratio and to the effective yield in clusters) since,
as we extensivley argue in the paper, the IMF (universal or not) is 
``bottom--light'' with respect to Salpeter both in clusters and in the
Solar Neighbourhood. The M/L ratio corresponding to a 10~Gyr old Kroupa SSP
is also indicated.

Comparing the effective iron yield in Fig.~\ref{fig:yield_cl}a
to the local
one  estimated in Eq.~\ref{eq:yieldFe_SV} (i.e.\ about solar), it is evident
that the observed yield in clusters is at the very least a factor of 2,
more likely a factor of 3--4 higher. If the cluster value is interpreted as
representative of a universal IMF --- and not of a different IMF than the 
local one --- this implies that 70\% of the metals ever produced by the Milky 
Way disc
must have escaped into the intergalactic medium, for the metal content in
the Solar Noighbourhood is 3--4 times lower than expected from the cluster 
yield.

We shall now discuss if an effective yield that is so much higher can be 
reconciled
in principle with a Kroupa/Scalo IMF, within present uncertainties in the 
stellar yields (the $p_Z(M)$ in Eq.~\ref{eq:yield}).
In our approach, the most uncertain iron yield from SN~II has been essentially
calibrated to reproduce the observed [$\alpha$/Fe] plateau in halo stars
(\S\ref{sect:metals_salp_ssp} and \S\ref{sect:Silicon}). Therefore, the 
relevant stellar yield uncertainty is in the oxygen and silicon yields.
Oxygen is a well understood element theoretically since it is produced in
hydrostatic nuclear burning stages, and stellar yields from
different authors agree within a factor of two; the uncertainty
mainly comes from the assumed mass loss, convection treatment and 
\Carbon$(\alpha,\gamma)$\Oxygen\ reaction rate \citep{GLM97,Pra98}. 
The silicon production is in principle more uncertain, being related to
explosive nucleosynthesis; however, from the analysis by \citet{GLM97} it 
seems that available SN~II yields in literature are in remarkable agreement
for \Silicon\ (within 20\%).

From Fig.~\ref{fig:yield_cl} we see that, for the stellar M/L ratio expected 
from the Kroupa IMF, the iron yield in clusters is 3--3.5 times larger than 
solar (and for $\alpha$--elements, the difference is up to a factor of 4--4.5;
panels c and d). This is a much larger discrepancy than the factor of two 
expected from existing sets of SN~II yields.
Therefore, at least to the extent that different theoretical predictions 
in literature are a fair measure of the real uncertainties in stellar yields, 
it seems that the metal enrichment observed in clusters cannot be reconciled 
with the local Kroupa/Scalo IMF via the uncertainties
in the SN~II yields alone. We even argued in \S\ref{sect:uncertainties}
that, judging from the large number of independent models of the Solar 
Neighbourhood
adopting different sets of yields, the actual uncertainty in metal production
related to the choice of stellar yields may be lower than that theoretically
affecting the SN models. In fact, in our conclusions 
(\S\ref{sect:conclusions}) we favoured either a variable IMF (different 
between clusters and disc galaxies) or a universal IMF that is 
different (in particular, shallower at the high mass end), than the
Kroupa/Scalo IMF.

The discrepancy in the observed effective yield between clusters and Solar 
Neighbourhood is even more drastic when $\alpha$--elements are considered;
and it is $\alpha$--elements, oxygen in particular, that define the bulk
of the metal mass.
Typical $\alpha$--element abundances in the stellar populations in clusters
are in the range [$\alpha$/H]=0 to +0.2~dex, or 
$Z_{\alpha, *} = 1-1.6 \, Z_{\alpha, \odot}$ 
(Case B and A in this paper, respectively). In the ICM [$\alpha$/Fe] abundance
ratios are at least solar \citep[e.g.\ the case of sulfur,][]{Baum03}, if not
significantly supersolar (e.g.\ the case of silicon). For oxygen in particular,
a solar abundance ratio [O/Fe]$\sim$0 seems to be appropriate if the updated 
solar iron abundance is adopted \citep{IA97}; which implies 
$Z_{O, ICM} = 0.34 \, Z_{O, \odot}$ from  the iron abundance determined by 
\citet{DeGra03} and adopted above in our calculations. 
We therefore compute the
effective yield of oxygen in cluster as:
\begin{equation}
\label{eq:yieldO_cl}
y_{O, cl} \sim \frac{(1-1.6 \, Z_{O, \odot}) \times M_* + 
0.34 \, Z_{O, \odot} \times M_{ICM}}{M_*}
\end{equation}
and plot it as a function of the assumed stellar M/L ratio (i.e.\ $M_{ICM}/M_*$
ratio) in Fig.~\ref{fig:yield_cl}c.
The oxygen yield in clusters is between 2.5 and 4 times larger than the local,
solar one; a factor of 4 discrepancy correspond to realistic
``bottom--light'' M/L ratios.

Oxygen is, however, not particularly well measured in the ICM; silicon is the
best measured element after iron \citep{Baum03}. For this element, ASCA data
provide SiMLR$\geq$0.01~\Msol/\Lsol\ \citep{F2000,Fin03, Loe2003}. For the 
stellar metallicity, we adopt $Z_{Si, *} = 1-1.6\, Z_{Si, \odot}$ as for the 
$\alpha$--elements
in general (see above). Hence we can compute the effective silicon yield 
in clusters as:
\begin{equation}
\label{eq:yieldSi_cl}
y_{Si, cl} \sim \frac{(1-1.6 \, Z_{Si, \odot}) \times M_* + M_{Si, ICM}}{M_*}
\end{equation}
where: 
\[ \frac{M_{Si, ICM}}{M_*} = \frac{SiMLR}{M_*/L_B} = \frac{0.01}{M_*/L_B} \]
and we plot the results in Fig.~\ref{fig:yield_cl}d.
The observed 
silicon yield in clusters is much larger, a factor of 3 to 4.5 --- with the
latter being more realistic --- than
the local observed solar yield. And we have been conservative in our 
calculations, since we adopted the lower limit of the observational range
SiMLR=0.01--0.02~\Msol/\Lsol\ \citep{F2000,Fin03}. In fact, the silicon 
abundance and [Si/Fe] ratios in the ICM are so large, that some contribution
from exotic Population~III hypernov\ae\ may be needed to produce it
\citep{Loe2001}.

All in all the effective yield in clusters is, with respect to the 
local solar one, about 3 times higher for iron and 4 or more times higher for
$\alpha$--elements. This difference is too large to be compensated by 
uncertainties in stellar nucleosynthesis alone, which are a factor of two or 
less.
If this discrepancy is not the result of a cluster IMF different from
the standard Solar Neighbourhood one, but it is representative of some 
universal IMF acting both in clusters and in the Milky Way, then
winds and  outflows must have ejected into the intergalactic medium 70\% 
of the metals actually produced in the Solar Neighbourhood. A challenge
indeed, for models of the evolution of disc galaxies.

\bibliographystyle{apjl}
\bibliography{biblio}

\end{document}